\documentclass[a4paper,12pt]{article}
\pdfoutput=1 

\usepackage{jheppub} 

\usepackage[T1]{fontenc} 
\usepackage{mathtools}

\usepackage{enumitem}

\usepackage{aas_macros}
\usepackage{amsmath,amssymb}
\usepackage{multirow}
\usepackage{cleveref}
\usepackage{siunitx}
\usepackage{enumerate}
\usepackage[dvipsnames]{xcolor}
\usepackage{float}

\usepackage[compat=1.1.0]{tikz-feynman}
\usepackage{afterpage}

\usepackage{amsthm}

\usepackage{psfrag}
\usepackage{color}
\usepackage{graphicx}
\usepackage{upgreek}
\usepackage{feynmp}
\DeclareMathAlphabet{\mathpzc}{OT1}{pzc}{m}{it}

\usepackage{hyperref}
\usepackage[mathscr]{euscript}

\usepackage[toc,page]{appendix}
\usepackage{circledsteps}
\usepackage{natbib}
\setcitestyle{square,comma,numbers,sort&compress}
\usepackage{enumerate}
\usepackage{tabularx,booktabs}
\usepackage[dvipsnames]{xcolor}
\usepackage{comment}
\usepackage[caption=false]{subfig}
\usepackage[compat=1.1.0]{tikz-feynman} 

\usepackage[section]{placeins}
\usepackage{cleveref}

\allowdisplaybreaks[4]

\definecolor{darkgreen}{rgb}{0,0.5,0}

\newcommand{\beq}{\begin{eqnarray}}
\newcommand{\eeq}{\end{eqnarray}}

\newcommand{\Tr}{{\rm Tr}}

\newcommand{\bseq}{\begin{subequations}}
\newcommand{\eseq}{\end{subequations}}
\newcommand{\be}{\begin{equation}}
\newcommand{\ee}{\end{equation}}

\renewcommand{\Im}{\mathop{\rm Im}\nolimits}

\newcommand{\eq}[1]{(\ref{#1})}

\newcommand{\beqa}{\begin{eqnarray}}
\newcommand{\eeqa}{\end{eqnarray}}

\usepackage{simpler-wick}
\newcolumntype{Y}{>{\centering\arraybackslash}X}

\newcommand{\vev}[1]{\langle #1 \rangle}

\title{Electric Dipole Moments as indirect probes of Dark Sectors}

\author{Marco Ardu,}
\author{Moinul Hossain Rahat,}
\author{Nicola Valori,} 
\author{and Oscar Vives}
\affiliation{Instituto de F\'isica Corpuscular, Universidad de Valencia and CSIC, Edificio Institutos Investigaci\'on, C/Catedr\'atico
Jos\'e Beltr\'an 2, 46980 Paterna, Spain}

\emailAdd{marco.ardu@ific.uv.es}
\emailAdd{moinul.rahat@ific.uv.es}
\emailAdd{nicola.valori@uv.es}
\emailAdd{oscar.vives@uv.es}

\abstract{
Dark sectors provide beyond Standard Model scenarios which can address unresolved puzzles, such as the observed dark matter abundance or the baryon asymmetry of the Universe. A naturally small portal to the dark sector is obtained if dark-sector interactions stem from a non-Abelian hidden gauge group that couples through kinetic mixing with the hypercharge boson. In this work, we investigate the phenomenology of such a portal of dimension five in the presence of CP violation, focusing on its signatures in fermion electric dipole moments. We show that, currently unbounded regions of the parameter space from dark photon searches can be indirectly probed with upcoming electron dipole moment experiments for dark boson masses in the range $1-100$ GeV. We also discuss two particular scenarios where a $SU(2)_D$ dark gauge group spontaneously breaks into either an Abelian $U(1)_D$ or nothing. In both cases, we show that potentially observable electron dipole moments can be produced in vast regions of the parameter space compatible with current experimental constraints and observed dark matter abundance. 
}

\begin{document}
\maketitle
\flushbottom
\section{Introduction}
The Standard Model (SM) of particle physics, despite its remarkable success in explaining a vast array of phenomena, remains incomplete. Crucially, it fails to account for Dark Matter (DM) and the Baryon Asymmetry of the Universe (BAU), highlighting the need for physics Beyond the Standard Model (BSM).

Hidden sectors offer rich scenarios for viable DM candidates~\cite{Pospelov:2007mp, Arkani-Hamed:2008hhe}, and may also introduce the necessary ingredients for baryogenesis~\cite{Haba:2010bm, Shelton:2010ta}. Given that SM interactions arise from local gauge symmetries, it is plausible that dark sector interactions could also stem from a hidden gauge group $G_D$. The simplest gauge theories involve an Abelian group $G_D = U(1)_D$~\cite{Chu:2011be}, but non-Abelian dark sectors have also attracted significant attention in DM model building~\cite{Buttazzo:2019iwr, Buttazzo:2019mvl, Landini:2020daq, Ardu:2020qmo, Frigerio:2022kyu, Borah:2022phw}. DM stability could result from accidental symmetries of the non-Abelian gauge theory, which may or may not be spontaneously broken, and DM could also be the bound state of a confining group \cite{ Soni:2016gzf, Yamanaka:2019yek}.

A natural candidate for a portal interaction connecting two gauge theories is the kinetic mixing between the gauge bosons \cite{Holdom:1985ag}. For an Abelian dark gauge group mixing with the SM hypercharge, the stringent experimental constraints require the coupling of the kinetic mixing $\kappa F^{\mu \nu} F'_{\mu \nu}$ to be small, even though such an operator is gauge-invariant and dimension-four. Intriguingly, for a non-Abelian dark sector, gauge invariance forbids the kinetic mixing at the renormalizable level, which facilitates a more natural explanation of the smallness of the interactions between the visible and the hidden sectors. Indeed, for a non-Abelian dark group, kinetic mixing can only arise from an effective operator of at least dimension five:
\begin{equation}
\frac{1}{\Lambda} \Sigma^a G_{\mu\nu}^a {B}^{\mu\nu}, \label{eq:CPeven}
\end{equation}
where $\Sigma^a$ is an adjoint scalar field, and $G^a_{\mu\nu}$, $B_{\mu\nu}$ are the dark and hypercharge gauge field strength tensors respectively. The above operator is possible if heavy states with masses $\sim \Lambda$ are charged under both hypercharge and the hidden gauge group\footnote{Higher-dimensional operators are possible choosing different scalar representations.}. Following the spontaneous breaking of the $G_D$ group, the scalar adjoint can acquire a vacuum expectation value (VEV), leading to the kinetic mixing between the SM and hidden sector gauge bosons. The phenomenology of this scenario has been extensively studied~\cite{KinMix0,KinMix1, KinMix2, KinMix3, KinMix4, KinMix5, KinMix6, KinMix7, KinMix8, KinMix9, KinMix10, KinMix11, KinMix12}. Alternatively, the non-Abelian group may remain unbroken and confine at low energy, with the above operator inducing kinetic mixing between a composite dark photon state and the SM hypercharge boson \cite{Alonso-Alvarez:2023rjq}.

If the hidden sector contains sources of CP violation, which may be necessary for baryogenesis, the CP-odd analogue of the operator in Eq.~(\ref{eq:CPeven}) is also possible:
\begin{equation}
	\frac{1}{\Lambda} \Sigma^a G_{\mu\nu}^a \tilde{B}^{\mu\nu},  \label{eq:CPodd}
\end{equation}
where $\tilde{B}^{\mu\nu}$ is the dual hypercharge tensor $\tilde{B}^{\mu\nu} = \epsilon^{\alpha \beta \mu \nu} B_{\alpha\beta} / 2$. The operators in Eqs.~(\ref{eq:CPeven}) and (\ref{eq:CPodd}) arise from similar UV diagrams and, in the presence of large CP-violating phases, have comparable coefficients. In this paper, we study the phenomenology of the CP-odd portal and discuss its complementarity with the more studied CP-even mixing. The phenomenology of a similar operator has been explored in the case of an Abelian dark photon mixing with the $SU(2)_L$ gauge bosons~\cite{Fuyuto:2019vfe, Cheng:2021qbl}. Here, we focus on the scenario where a non-Abelian dark gauge group kinetically mixes with the hypercharge.

Notably, Electric Dipole Moments (EDMs) provide the most stringent tests of CP-violating new physics. Assuming that the hidden sector group $G_D$ is spontaneously broken by the adjoint scalar VEV, the operator of Eq.~(\ref{eq:CPodd}) contributes to the EDM of SM fermions under very general assumptions. Given the impressive experimental sensitivity of the electron EDM (eEDM) searches~\cite{Roussy:2022cmp}, which is also expected to improve further in the future~\cite{ACMEIII_Hiramoto_2023, YBFFitch_2021, BaFPhysRevA.98.032513}, the eEDM arising from the operator of Eq.~(\ref{eq:CPodd}) could be detectable. In this work, we assume the presence of such an operator and investigate the potential of eEDM searches in probing a wide variety of dark sectors. 

The paper is organized as follows. In Section \ref{sec:nonabelianmix} we discuss the main features of the non-Abelian kinetic mixing and calculate, under some general assumptions, the contribution to the electron EDM coming from the CP-odd portal. We also explore the interplay between the dark photon searches and the eEDM experiments in this context.  In Section \ref{sec:DSmodels} we examine the specific case of a $SU(2)_D$ group broken into either an Abelian sub-group $U(1)_D$ or completely. In the former scenario, we find that the dark sector particles acquire milli-charges that can be indirectly probed using EDMs. In the case of a completely broken $SU(2)_D$ we implement an inelastic DM model, showing that large eEDMs are predicted across a vast region of the parameter space compatible with the observed DM abundance. Additional details regarding model building are relegated to the Appendices \ref{app:UV}, \ref{app:barr-zee} and \ref{app:iDM}. Finally, in Section \ref{sec:concl} we summarize our results. 

\section{Non-Abelian kinetic mixing}\label{sec:nonabelianmix}
We consider a hidden sector that is invariant under a non-Abelian gauge group \(G_D\), with SM particles transforming as singlets. We assume that \(G_D\) is spontaneously broken by the dark scalar sector, which contains at least one scalar \(\Sigma\) transforming in the adjoint representation of $G_D$. The Lagrangian takes the following general form:
\begin{equation}\label{eq.2.1}
\mathcal{L} = \mathcal{L}_{\rm SM} + \mathcal{L}_{\rm DS} + \text{Tr}[(D_\mu \Sigma)^\dagger(D^\mu \Sigma)] - V(\Sigma) + \mathcal{L}_{\rm SM-DS},
\end{equation}
where \(\mathcal{L}_{\rm SM}\) is the SM Lagrangian, \(\mathcal{L}_{\rm DS}\) contains the dark sector particles and their interactions, \(V(\Sigma)\) is the potential of the adjoint scalar \(\Sigma\), and \(\mathcal{L}_{\rm SM-DS}\) contains the interactions that connect the visible and hidden sectors. In the presence of heavy states that have hypercharge and are also charged under $G_D$, mixing between the SM and dark gauge bosons can arise at dimension five via the operators
\begin{equation}
\mathcal{L}_{\rm SM-DS} \supset -\frac{C}{\Lambda} \text{Tr}[\Sigma X^{\mu \nu}] B_{\mu \nu} - \frac{\tilde{C}}{\Lambda} \text{Tr}[\Sigma X^{\mu \nu}] \tilde{B}_{\mu \nu},\label{eq:nonabelianmixingops}
\end{equation}
where \(X^{\mu \nu} = T_{a} X_{a}^{\mu \nu}\) is the \(G_D\) non-Abelian field strength tensor, defined as \(X_{a}^{\mu \nu} = \partial^{\mu} X_{a}^{\nu} - \partial^{\nu} X_{a}^{\mu} + g_D f_{abc} X_{b}^{\mu} X_{c}^{\nu}\). Here, \(T^{a}\) are the \(G_D\) generators, \(g_D\) is the hidden sector gauge coupling, and \(f_{abc}\) are the structure constants of the \(G_D\) group. $\Lambda$ is the UV cut-off and is related to the mass scale of the heavy states that are integrated out to generate the operators above (see Appendix \ref{app:UV} for a possible UV completion). When both terms of Eq~(\ref{eq:nonabelianmixingops}) are present in the Lagrangian, CP is broken because the contractions $X^{\mu\nu} B_{\mu\nu}$ and $X^{\mu\nu} \tilde{B}_{\mu\nu}$ are respectively even and odd under CP.  

If the  scalar field $\Sigma$ acquires a VEV $\langle \Sigma \rangle=v_aT^a$, upon spontaneous symmetry breaking, Eq.~(\ref{eq:nonabelianmixingops}) becomes
  \begin{equation}
	\mathcal{L}_{\rm SM-DS} \supset - \frac{\epsilon_{a}}{2}  X_{a}^{\mu \nu} B_{\mu \nu} - \frac{\epsilon_a}{2 v_a} \phi^{a} X_{a}^{\mu \nu} B_{\mu \nu}- \frac{\tilde{\epsilon}_a}{2} X_{a}^{\mu \nu} \tilde{B}_{\mu \nu}- \frac{\tilde{\epsilon}_a}{2v_a}\phi^{a} X_{a}^{\mu \nu} \tilde{B}_{\mu \nu},\label{eq:SSBops}
  \end{equation}
 where $\phi^a \equiv \Sigma^a-v_a$, we have assumed that the $G_D$ generators are normalized as $\Tr(T^aT^b)=\delta^{ab}/2$, and we have defined $\epsilon_a=Cv_a/\Lambda$ and $\tilde{\epsilon}_a=\tilde{C}v_a/\Lambda$.

\subsection{Gauge sector}
The gauge boson quadratic terms, following the spontaneous breaking of the electroweak $SU(2)_L\times U(1)_Y$ and dark gauge groups, are 
\begin{eqnarray}
    \mathcal{L}_{\rm gauge}&\supset& - \frac{1}{4} F^{\mu \nu}F_{\mu \nu} - \frac{1}{4} Z^{\mu \nu}Z_{\mu \nu}- \frac{1}{4} X^{\mu \nu}_{a}X_{\mu \nu}^{a} +\frac{1}{2}m_{Z}^{2}Z^{\mu}Z_{\mu}+\frac{1}{2}M^{2}_{ab} X^{a}_{\mu}X^{b\mu}\nonumber\\
    &-& \frac{\epsilon_{a}}{2}c_{\theta} X_{a}^{\mu \nu} F_{\mu \nu}  + \frac{\epsilon_{a}}{2}s_{\theta} X_{a}^{\mu \nu} Z_{\mu \nu},
\label{eq:mixedbilinear}
\end{eqnarray}
where we have added the kinetic mixing terms coming from Eq.~(\ref{eq:SSBops}), and $c_{\theta}$ and $s_{\theta}$ are the cosine and sine of the Weinberg angle. The CP-odd mixing term $\tilde{\epsilon}_a X_{a}^{\mu \nu} \tilde{B}_{\mu \nu}$ is not included because its contribution to the two-point function is a total derivative. 

Although a complete description of the gauge boson mixing would require specifying the SSB of the dark gauge symmetry in detail, to illustrate the pheonomenological consequence of the effective operators we consider only one dark gauge boson $X$ with  a mass $M_X$ that mixes with the $B$ field with a kinetic mixing parameter $\epsilon$. This leads to the following rotation matrix

\begin{equation}\label{rotgen}
	\begin{pmatrix}
		A \\
		Z \\
		X 
	\end{pmatrix}=\begin{pmatrix}
		1 & 0 & -c_{\theta} \epsilon  \\
		0& 1 &  -\frac{s_{\theta} \epsilon M_{X}^{2}}{M_{Z}^{2}-M_{X}^{2}} \\
		0  & \frac{s_{\theta} \epsilon M_{Z}^{2}}{M_{Z}^{2}-M_{X}^{2}}  &  1\\
	\end{pmatrix}\begin{pmatrix}
	A' \\
	Z'\\
	X'
	\end{pmatrix},
\end{equation}
where the primed fields correspond to the mass eigenstates, as can be seen if we replace this transformation in Eq.~(\ref{eq:mixedbilinear}),  and $M_Z$ is the $Z$ boson mass. As a result of the rotation, the dark gauge boson couples to the electromagnetic and $Z$ currents with an $\epsilon$-suppressed coupling,
\begin{equation}
 \mathcal{L}_{\rm gauge}\supset -X'_\mu (c_\theta \epsilon e J^\mu_{\rm EM}+ \epsilon e \tan \xi J^\mu_Z),
\end{equation}
where $\tan \xi \equiv M^2_X/(M_Z^2-M_X^2)$ and the $Z$ current is defined as $J^\mu_Z \equiv (J^\mu_3 -s_\theta^2J^\mu_{\rm EM})/c_\theta$, with $J^\mu_3$ being the current of the $T_3$ generator of $SU(2)_L$. In addition, the physical $Z$ boson state can couple to the dark sector particles via an $\epsilon$-dependent coupling, which can be probed by electroweak precision observables and rare $Z$ decays \cite{Hook:2010tw, ATLAS:2023jyp}.

\subsection{Scalar sector}
The scalar sector responsible for the spontaneous breaking of the dark gauge group depends on the specifics of the hidden sector. We will discuss some particular models in Section \ref{sec:DSmodels}, but for now we assume the presence of a single scalar transforming as an adjoint of $G_D$, which can interact with the SM Higgs via the quartic-coupling
\begin{equation}
	\mathcal{L}_{\rm SM-DS}\supset \lambda(H^\dagger H)\Tr(\Sigma^\dagger \Sigma).
\end{equation}

Once the Higgs and dark scalar acquire VEVs, the quartic portal induces  mixing between the Higgs field ($h \equiv H - \vev{H}$) and the dark scalars ($\phi_{a} \equiv \Sigma_a - \vev{\Sigma_a}$). While in general several scalar states could mix, we simplify the analysis by considering the case where only one scalar, $\phi$, mixes significantly with the physical Higgs, and focus on the sub-matrix
\begin{equation}\label{Higgsrotation}
\begin{pmatrix}
	h \\
	\phi 
\end{pmatrix}\simeq 
\begin{pmatrix}
    1 & -\beta  \\
    \beta & 1  \\
\end{pmatrix}
\begin{pmatrix}
h' \\
\phi'
\end{pmatrix}.
\end{equation}
Here, the primed fields are mass eigenstates, and we expanded at leading order in the mixing angle $\beta$, which is constrained to be small by collider searches \cite{Ferber_2024}. If $\phi'$ is heavier than $m_{h}$/2,  ATLAS \cite{ATLAS2022} and CMS \cite{CMS:2022dwd} constrain the mixing of a singlet scalar with the Higgs to be $\sin\beta$ < 0.27. If instead $\phi'$ is light enough to allow the decay of the SM Higgs into a pair of new singlet scalars,  a stronger bound ($\beta \lesssim 10^{-2}$) is determined by the searches of the Higgs invisible decay branching ratio \cite{Atlas22022, CMS2PhysRevD.105.092007}. 

\subsection{Electric dipole moments from the CP-odd kinetic mixing}
\renewcommand{\arraystretch}{1}
After rotating the fields in the mass eigenstate basis, the Lagrangian contains the following interactions\footnote{For notational convenience we drop the prime in the mass eigenstate fields in the following.}
\begin{eqnarray}
	\mathcal{L}\supset &-& \frac{\tilde{\epsilon}}{2v_D}(\phi+\beta h) \left(c_\theta X^{\mu \nu} \tilde{F}_{\mu\nu} -s_\theta X^{\mu \nu} \tilde{Z}_{\mu\nu} +c_\theta s_\theta\frac{\epsilon M_Z^2}{M_Z^2-M^2_X}Z^{\mu\nu} \tilde{F}_{\mu\nu}\right)\nonumber\\
	&-&\sum_f \left(Y_f  \frac{h}{\sqrt{2}} \bar{f} f-Y_f \beta \frac{\phi}{\sqrt{2}} \bar{f} f\right)
	-X_\mu (c_\theta \epsilon e J^\mu_{\rm EM}+ \epsilon e \tan \xi J^\mu_Z), \label{eq:intEDM}
\end{eqnarray}
where $v_D$ is the dark scalar VEV and the sum over $f$ run over all SM fermions with a Yukawa coupling $Y_f$. The terms in the first line of Eq.~(\ref{eq:intEDM}) are CP-odd and can contribute to CP-violating observables.  
\begin{figure}[t]
	\centering
	\includegraphics[width = 0.4\linewidth]{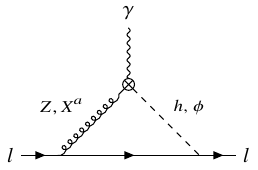}
	\caption{One-loop contribution to fermion dipole moments.}
	\label{fig_feynman_EDM}
\end{figure}
For instance, 
the diagram in Fig. \ref{fig_feynman_EDM} illustrates the contribution to fermion EDMs that stem from the above interactions. Similar diagrams can be drawn with the scalar propagator replaced by a dark gauge boson, arising from the vertex $X^a_{\mu\nu}\tilde{F}^{\mu\nu}\supset g_D f^{abc}X^b_{\mu}X^c_{\nu}\tilde{F}^{\mu\nu}$, but their contribution vanishes due to the antisymmetry of $f^{abc}$ when summing over all possible diagrams.

In agreement with Furry's theorem \cite{PhysRev.51.125}, we find that the diagrams in Fig.~\ref{fig_feynman_EDM} select the vector component of the fermion current at the gauge boson vertex. As a consequence of this selection rule, the contributions to the lepton EDMs arising from the $Z$ boson diagram and from the $X$ coupling to the lepton $Z$ current are suppressed,
because the $Z$ lepton vector current is proportional to the accidentally small combination $1-4 s_\theta^2\sim 0.08$.  As a result, the predicted lepton EDM can be approximated as
\begin{eqnarray}
	d_{l} &\simeq& \frac{Y_{l} }{8 \pi^{2}v_D} \epsilon \tilde{\epsilon} \beta c_{\theta}^{2}\, e\ \left(\frac{\log(x_{X\phi})}{x_{X\phi}-1}-\frac{\log(x_{Xh})}{x_{Xh}-1}\right)\nonumber\\
	&\equiv&\frac{Y_{l} }{8 \pi^{2}v_D} \epsilon^2 \tan\chi\ \beta c_{\theta}^{2}\, e\ \left(\frac{\log(x_{X\phi})}{x_{X\phi}-1}-\frac{\log(x_{Xh})}{x_{Xh}-1}\right),\label{eq:EDM}
\end{eqnarray}
where $x_{ij}=m^2_{i}/m^2_{j}$ and in the second line we have defined $\tan \chi \equiv \tilde{\epsilon}/\epsilon$, which in the UV is related to the phase of the Yukawa couplings of the heavy states that are integrated out to give the effective operators (see Appendix \ref{app:UV}). Values $\tan \chi \sim \mathcal{O}(1)$ correspond to large CP-violating phases in the UV, which will be our assumption here and in the rest of the paper. 

EDMs are known to be extremely sensitive probes of CP-violating new physics.
In particular, the electron EDM provides the most stringent upper limit on the couplings of Eq.~(\ref{eq:EDM}), due to the impressive sensitivity of current experimental searches, which are also expected to improve significantly in the future. We summarize the current and future status of electron EDM searches in Table \ref{tab:bds}. Efforts are underway to improve the experimental sensitivity to the muon EDM at PSI \cite{Sakurai:2022tbk}, but the expected reach is not able to constrain the parameter space more than the electron EDM in this scenario. This is because in our model we do not consider new flavor-dependent couplings beyond the SM Yukawa couplings, and therefore the contributions to the EDMs follow the Minimal Flavour Violation (MFV) expectation, where the Yukawa couplings are responsible for the chirality flips. Hence, the scaling of the lepton EDM is such that the indirect bound on the muon EDM arising from the electron searches is orders of magnitude stronger than the expected muon EDM sensitivity.

\renewcommand{\arraystretch}{1.3}
\begin{table}[t]
	\begin{center}
		\begin{tabular}{c c}
			\toprule
			Experiment & Current bound/Upcoming sensitivity    \\
			\midrule
			JILA eEDM &\qquad \qquad $<4.1\times 10^{-30}$ ${\rm e \ cm^{-1}}$  \cite{Roussy:2022cmp}\\
			ACME III &\qquad \qquad $\sim1\times 10^{-30}$ ${\rm e \ cm^{-1}}$  \cite{ACMEIII_Hiramoto_2023}\\
			YBF &\qquad \qquad $\sim1\times 10^{-31}$ ${\rm e \ cm^{-1}}$  \cite{YBFFitch_2021}\\
			BaF &\qquad \qquad $\sim1\times 10^{-33}$ ${\rm e \ cm^{-1}}$  \cite{BaFPhysRevA.98.032513}\\
			
			\bottomrule
	\end{tabular}	\end{center} 
	\caption{Current (JILA eEDM) bound on the electron EDM and expected future sensitivities of the upcoming searches.
		\label{tab:bds}} 
\end{table}

\begin{figure}[ht]
	\centering
	\includegraphics[width = 0.8\linewidth]{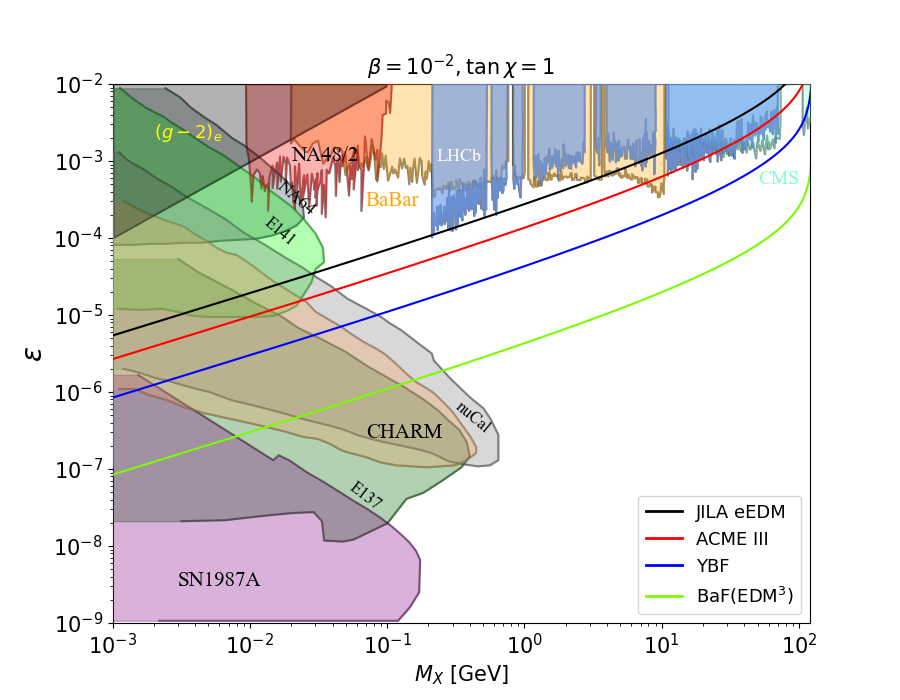}
	\caption{Plot of kinetic mixing parameter $\epsilon$ vs dark boson mass. Colored bands represent the region of parameter space saturated by the current (JILA eEDM \cite{Roussy:2022cmp}) and future (ACME III \cite{ACMEIII_Hiramoto_2023}, YBF \cite{YBFFitch_2021}, BaF($\mathrm{EDM}^{3}$)\cite{BaFPhysRevA.98.032513}) experimental sensitivity of the eEDM assuming fixed scalar mixing angle $\beta = 10^{-2}$ and a large CP-violating phase tan$\chi$ = 1. Bounds on kinetic mixing parameter $\epsilon$ come from di-lepton final states searches at colliders (LHCb \cite{LHCb_ZbosoPhysRevLett.124.041801}, BaBar \cite{BaBar_Zboson:2014zli}, CMS \cite{CMS_Zboson:2019kiy} and NA48/2 \cite{NA482_Zboson:2015wmo}) and beam dump experiments (E137 \cite{E137_Zboson_Bjorken:1988as,E137_Zboson_Batell:2014mga,E137_Zboson_Marsicano_2018}, E141\cite{E141_ZbosonRiordan:1987aw}, CHARM \cite{CHARM_Zboson_Gninenko_2012}, $\nu$-Cal \cite{nucal_Zboson_Bl_mlein_2014}, NA64 \cite{NA64_Zboson_Banerjee_2018}). Constraints from SN1987A \cite{Supernova_Zboson_Chang_2017} and $g-2_e$ \cite{g_2e_Zboson_Pospelov_2009} are also shown. }
	\label{kinetic_mixing_plot}
\end{figure}

To reduce the number of free parameters in Eq.~(\ref{eq:EDM}), we assume that the dark sector couplings are natural, with only one mass scale set by the scalar VEV \(v_D\). Thus, we take \(v_D \sim M_X \sim m_\phi\) and study the EDM predictions by varying the other parameters. We focus on the region \(1\ \text{GeV} \lesssim M_X \lesssim 100\ \text{GeV}\) because lighter extra gauge bosons that mix with the photon are severely constrained by a combination of beam dump experiments and astrophysical probes 
\cite{E137_Zboson_Bjorken:1988as,E137_Zboson_Batell:2014mga,E137_Zboson_Marsicano_2018, E141_ZbosonRiordan:1987aw, CHARM_Zboson_Gninenko_2012, nucal_Zboson_Bl_mlein_2014, Supernova_Zboson_Chang_2017}. Fig.~\ref{kinetic_mixing_plot} shows values of \(\epsilon\) and \(M_X\) that saturate the present and future electron EDM sensitivities shown in Table~\ref{tab:bds}, having chosen a scalar mixing angle \(\beta\) at the boundary of what is allowed by Higgs invisible decays. We stress that Fig.~\ref{kinetic_mixing_plot} shows the current and future sensitivity of the EDM to the kinetic mixing parameter \(\epsilon\), assuming it is generated by a non-renormalizable operator that connects a non-Abelian dark sector with the SM  and in the presence of CP-violating phases that generate a  CP-odd operator with a comparable coefficient. In this scenario, the EDM can probe a large region of the parameter space in \(\epsilon\) and dark gauge boson mass \(M_X\) that is currently unexplored by the dark photon searches \cite{LHCb_ZbosoPhysRevLett.124.041801, CMS_Zboson:2019kiy, BaBar_Zboson:2014zli, NA482_Zboson:2015wmo}. In Section \ref{sec:DSmodels}, we discuss how this can be used to probe specific dark sector models.

As an aside, the model would contribute to the electron $g-2$ ($a_{e}$) through the CP-even analogue of the diagram in Fig.~\ref{fig_feynman_EDM}, and the final result will resemble Eq.~\eqref{eq:EDM} up to a factor 2$m_{e}$ and $\tilde{\epsilon} \rightarrow \epsilon$. However, considering the most precise measurement of $a_{e}$ \cite{eg-2Morel:2020dww}, it will give a milder constraint on the kinetic mixing $\epsilon$ within our assumptions. 

\section{Models}\label{sec:DSmodels}
In this section we discuss some dark sector scenarios that could be probed with the electron EDM induced by the CP-odd portal introduced in the previous section.
\subsection{$SU(2)_D\to U(1)_D$} \label{model1}
The minimal setup that satisfy the assumptions of Section \ref{sec:nonabelianmix} is given  by a $SU(2)_D$ group which is broken by a triplet into an Abelian sub-group $U(1)_D$. The Lagrangian in this case is given by
\begin{equation}
		\mathcal{L} = \mathcal{L}_{\rm SM} + \mathcal{L}_{\rm DS} + (D_\mu \Sigma^a)^2 - V(\Sigma, H) -\frac{\epsilon}{2 v_D} \Sigma^a X^{a\mu \nu} B_{\mu \nu} - \frac{\tilde{\epsilon}}{2 v_D} \Sigma^a X^{a\mu \nu} \tilde{B}_{\mu \nu},
\end{equation}
where $a$ takes values from 1 to 3, and the covariant derivative acting on the triplet is defined as $D_\mu\Sigma^a= \partial_\mu \Sigma^a+\epsilon^{abc} \Sigma^b X_\mu^c$. The effective operators arise from integrating out heavy fermions that have hypercharge and fill non-trivial representations of $SU(2)_D$ (see Appendix \ref{app:UV}). The operator normalization $v_D$ is the triplet VEV, which is given by the minimization of the $V(\Sigma, H)$ potential
\begin{align}
   V(\Sigma, H) &=-\mu^2_H (H^\dagger H)+\lambda_H(H^\dagger H)^2 -m^2_\Sigma (\Sigma^a\Sigma^a) +\lambda_{\Sigma} (\Sigma^a\Sigma^a)^2 \nonumber \\
   &+\lambda_{H\Sigma}(H^\dagger H)(\Sigma^a \Sigma^a).\label{eq:scalarpotentialmassless}
\end{align}

Without loss of generality, we assume that the triplet vacuum is aligned with the direction of the diagonal generator of the $SU(2)_{D}$ algebra, with  $\langle \Sigma^3 \rangle=v_D$ and $\langle \Sigma^{1,2} \rangle=0$. Consequently, the $X^{1,2}_\mu$ boson will acquire a squared mass $M^2_X=g_D v^2_D$, while $X^3_\mu$ remains massless and has a non-zero kinetic mixing with the hypercharge boson.  When the electroweak symmetry is spontaneously broken, the massless vector will mix with the photon and the $Z$ boson. Since two massless vector fields mix kinetically, there is an ambiguity in identifying the physical states, one of which should be the SM photon. The ambiguity arises because the kinetic terms are invariant under an arbitrary orthogonal transformation, which leaves the physics unchanged \cite{PAN2020114968}. To correctly interpret the model predictions, we need to specify how the extra massless boson interacts with the SM fields. If we define the physical dark state as the combination that interacts only with the hidden sector particles, we find the following mixing matrix in the gauge sector
\begin{equation}\label{rotgaugemassless}
	\begin{pmatrix}
        A' \\
		Z'\\
		X_3
	\end{pmatrix}=\begin{pmatrix}
		1 & 0 & 0  \\
		0& 1 &  0 \\
	  -c_{\theta} \epsilon & s_{\theta}\epsilon  &  1\\
	\end{pmatrix}\begin{pmatrix}
		A \\
		Z\\
		A_D 
	\end{pmatrix},
\end{equation}
where now the unprimed fields are the mass eingenstates. As a result, the massless dark photon only interacts with the $U(1)_D$ charged particles, while the photon and the $Z$ boson couples to the SM, and to the hidden sector via a $\epsilon-$dependent coupling. In practice, this means that the $U(1)_D$ current is milli-charged
\begin{equation}
	\mathcal{L}\supset -e A_\mu \left( J^\mu_{\rm EM}+c_\theta \frac{\epsilon g_D}{e} J^\mu_{DS}  \right). \label{eq:millicharge}
\end{equation}
The triplet component that acquires a VEV can mix with the SM Higgs via the quartic coupling, as described in Eq.~(\ref{Higgsrotation}), with a mixing angle $\beta$ given by the Lagrangian parameter introduced in the scalar potential of Eq.~(\ref{eq:scalarpotentialmassless}). Rotating the gauge bosons and scalars into the physical states, we find the following interactions\footnote{The Lagrangian also contains the CP-even contraction $\sim \beta \epsilon^2/(2v_D) h F^{\mu\nu} F_{\mu\nu}$ that can interfere with the SM amplitude for $h\to \gamma \gamma$. Since we assume that $\epsilon\sim \tilde{\epsilon}$ we find that the sensitivity of $h\to \gamma \gamma$ to $\epsilon^2/v_D$ does not compete with the EDM bounds, in agreement with the results of \cite{McKeen_2012}.} 
\begin{eqnarray}
	 \mathcal{L}&\supset &-\frac{\tilde{\epsilon}}{2v_D}(\phi+\beta h) \left(c^2_\theta \epsilon F^{\mu \nu} \tilde{F}_{\mu\nu} +s^2_\theta \epsilon Z^{\mu \nu} \tilde{F}_{\mu\nu} \right)\nonumber\\
	&-&\sum_f \left(-Y_f \beta \frac{\phi}{\sqrt{2}} \bar{f} f+Y_f  \frac{h}{\sqrt{2}} \bar{f} f\right),
\end{eqnarray}
where we have denoted $\phi$ as the mass eigenstate that results from the $\Sigma^3$ and Higgs mixing. The lepton EDM induced by the above interactions is given by a diagram similar to the one in Fig. \ref{fig_feynman_EDM}, but with 
$X$ replaced by the SM photon. The predicted lepton EDM  is
\begin{equation}
	d_{l}\simeq e \left(\frac{Y_{l}}{8 \pi^{2} v_D}\right)\epsilon^2\tan\chi \; \beta c_{\theta}^{2}\log \left(\frac{m_h^2}{m_\phi^2}\right),
\end{equation}
 where, again, we have neglected the $Z$ contribution because the lepton vector current coupling is suppressed. We remind that $\tan\chi$ is defined as $\tan\chi\equiv \tilde{\epsilon}/\epsilon$.
 
\begin{figure}[ht]
	\centering
	\includegraphics[width = 0.8\linewidth]{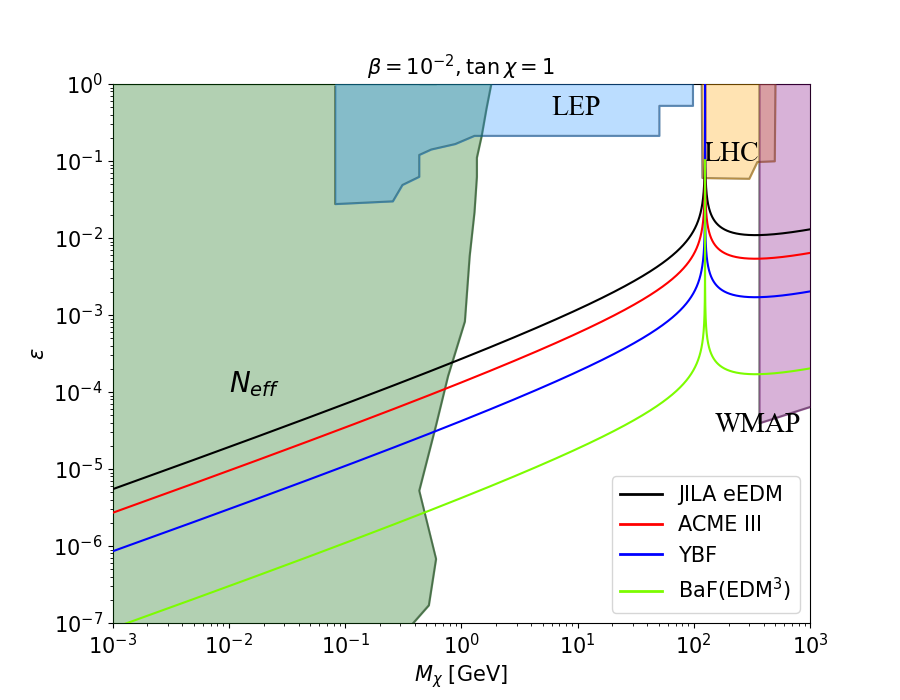}
	\caption{Plot of kinetic mixing parameter $\epsilon$ vs milli-charged particle mass. Colored bands represent the region of parameter space saturated by the current (JILA eEDM \cite{Roussy:2022cmp}) and future (ACME III \cite{ACMEIII_Hiramoto_2023}, YBF \cite{YBFFitch_2021}, BaF($\mathrm{EDM}^{3}$) \cite{BaFPhysRevA.98.032513}) experimental sensitivity of the eEDM assuming scalar mixing angle $\beta = 10^{-2}$, dark scalar mass $m_{\phi}\sim v_D$  and a large CP-violating phase tan$\chi$ = 1. Bounds on kinetic mixing parameter $\epsilon$ come from $N_{\mathrm{eff}}$ \cite{mQ1Vogel_2014}, LEP \cite{mQ6Davidson_2000}, LHC \cite{mQ7Jaeckel_2013}, WMAP \cite{WMAP_Dubovsky_2004}. In this plot, $g_{D} = e$ is assumed.}
	\label{millicharge_plot}
\end{figure}

 The dark sector Lagrangian $\mathcal{L}_{\rm DS}$ contains matter fields that transform under the $SU(2)_D$ group, and may include potential dark matter candidates.  After the SSB $SU(2)_D\to U(1)_D$, particles that are charged under the surviving $U(1)_D$ also have a milli-charge (mC)  $\epsilon'\equiv c_\theta \epsilon g_D/e $ as described by Eq.~(\ref{eq:millicharge}). Constraints on mC Particles (mCPs) come from a variety of experimental probes \cite{mQ6Davidson_2000, mQ1Vogel_2014,mQ2Chang_2018,mQ3Badertscher_2007,mQ4Prinz_1998,mQ5Magill_2019,mQ7Jaeckel_2013, mQ8PhysRevD.43.2314,Kalliokoski:2023cgw,WMAP_Dubovsky_2004}. In our scenario, the mC parametrizes the EDM prediction, so the EDM bounds are translated into an indirect sensitivity to the millicharge.  The $X_{1,2}$ massive bosons combine to form the  $U(1)_D$ charge eigenstates $W^{\pm}_D=(X_1\mp iX_2)/\sqrt{2}$, which interact with the SM photon via the mC. Assuming that the dark gauge coupling is comparable to the electromagnetic one, $g_D\sim e$, the EDM prediction can be parametrised in terms of the kinetic mixing $\epsilon$ and the $W_D$ mass\footnote{The EDM depends only logarithmically on the scalar mass, which we assume to be $m_\phi \sim v_D$}. The EDM sensivity can then be compared with the experimental reach of milli-charged particles searches (see Fig. \ref{millicharge_plot}). Clearly, other mC states besides the $W_D$ could be present in the hidden sector. As shown in Fig.~\ref{millicharge_plot}, the constraints on the mC can be severe enough to render the EDM unobservable for a wide range of masses, while interesting EDM phenomenology can arise if the mCP masses lies within $1\ {\rm GeV}\lesssim M_\chi\lesssim 100$ GeV
 
The millicharged particles can contribute to the dark matter relic density if they are stable. However, strong constraints from CMB observations either require a mC small enough to have dark matter decoupled before the recombination epoch, or that the mCPs constitute a small fraction of the observed dark matter abundance \cite{McDermott:2010pa}. In the first case, where the millicharged particles account for the dark matter but the millicharge is small  ($\epsilon' \lesssim 10^{-7}$ ), the EDM will also be suppressed, making it unobservable even for the most optimistic upcoming sensitivity. {However, UV-dependent contributions to EDMs arising from the integrated-out heavy fermions could still be sizable despite the small $\epsilon$-values, as we illustrate in Appendix \ref{app:barr-zee}.} If instead the mCPs constitute only a negligible fraction of the dark matter, larger $\epsilon'$
values are allowed and, in our scenario, an electron EDM could be observed for multi-GeV mCP masses. Interestingly, an anomalous observation of the 21 cm hydrogen line at red-shift $z\sim 17$ \cite{Bowman:2018yin} from the EDGES collaboration, could be explained by a small fraction of millicharged DM \cite{Liu:2019knx, Aboubrahim:2021ohe, Munoz:2018jwq, Mathur:2021gej}.

Viable DM candidates in this model must be stable particles uncharged under the surviving $U(1)_D$ to avoid the stringent bounds on mC dark matter. Since $U(1)_D$ charged and neutral states co-exist in $SU(2)_D$ multiples, accommodating a negligible mC DM fraction while simultaneously accounting for the observed DM abundance with the neutral components is challenging and beyond the scope of this work. We aim to address this issue in a future project.

{Notice that the minimal setup that we considered in this section for the dark sector, an $SU(2)_D$ symmetry with an scalar in the adjoint spontaneously broken to $U(1)_D$, is exactly the original magnetic monopole construction described by ’t Hooft and Polyakov \cite{tHooft:1974kcl,Polyakov:1974ek}. Therefore, we should expect dark sector monopoles. In \cite{Brummer:2009cs} it was shown that, in the presence of only kinetic mixing, these dark monopoles are invisible to ordinary electrically charged matter. However, in the presence of both the ordinary kinetic mixing and the CP-odd mixing, the dark monopoles acquire an electric charge in the dark sector through the Witten
effect \cite{Witten:1979ey}. Then, these "dark dyons" appear as mCPs in the visible sector, through the usual kinetic mixing \cite{Brummer:2009oul}\footnote{Constraints on dyons have been obtained in the MoEDAL experiment \cite{MoEDAL:2020pyb} but would not apply to these "dark dyons" with no magnetic charge in the visible sector.}. Nonetheless, for a monopole mass $\lesssim 1$~TeV like in our case, their contribution to the dark matter relic density is completely negligible, so cosmological and astrophysical bounds on the monopole millicharge are not severe. }

\subsection{$SU(2)_D\to \varnothing$ and inelastic Dark Matter} \label{model2}
Augmenting the scalar sector of Section \ref{model1} with an additional triplet, it is possible to completely break the $SU(2)_D$ group if the triplet VEVs are not aligned. Considering two scalar $\Sigma^a_1,\Sigma^a_2$ transforming in the adjoint of $SU(2)_D$, the potential can be such that
\begin{equation}\label{vacuumexpvalue}
	\langle \Sigma_1 \rangle=\begin{pmatrix}0 & v_1 & 0\end{pmatrix},\qquad \langle \Sigma_2 \rangle=\begin{pmatrix}0 & 0 & v_2\end{pmatrix},
\end{equation}
which results in the following mass matrix for the dark gauge bosons
\begin{equation}
	\mathcal{L}\supset \frac{1}{2} M^2_{ab} X^a_\mu X^{b\mu},\qquad M^2_{ab}=\begin{pmatrix}
		g_D (v^2_1+v^2_2) & 0 & 0\\
		0 & g_D v^2_2 & 0\\
		0 & 0 & g_D v_1^2
	\end{pmatrix},\label{eq:gbmasses}
\end{equation}
and the $X^2,X^3$ bosons acquire non-zero kinetic mixing with the SM gauge fields via the non-renormalizable portals
\begin{equation}
	\mathcal{L}\supset \sum_{i=1,2} \left(-\frac{\epsilon_i}{2 v_i} \Sigma_i^a X^{a\mu \nu} B_{\mu \nu} - \frac{\tilde{\epsilon}_i}{2 v_i} \Sigma_i^a X^{a\mu \nu} \tilde{B}_{\mu \nu}\right).
\end{equation}

In this case, the mass eigenstates aligned with $X_2,\ X_3$ interact with the electromagnetic current after the rotation that diagonalises the kinetic terms 
\begin{equation}
	\mathcal{L}\supset \sum_{i=1,2} X^{i+1}_\mu c_\theta \epsilon_i e J^{\mu}_{\rm EM}.
 \label{eq:Xemcoupl}
\end{equation}
The scalar sector features four physical states, one being the SM Higgs. Two of the scalar states, aligned with the two VEVs, $\Sigma^{2,3}_{1,2}=v_{1,2}+\phi^{1,2}$, will mix directly with the Higgs via the quartic coupling. Assuming for simplicity small mixing angles, we have the following matrix that rotates into the mass eigenstates basis
\begin{equation}
	\begin{pmatrix}
         \phi_D\\
		 \phi_1 \\
		 \phi_2 \\
		 h
	\end{pmatrix}=\begin{pmatrix}
 1 & \theta_1 & \theta_2 & 0\\
	-\theta_1&	1 & \alpha& \beta_1  \\
	-\theta_2&	-\alpha & 1 &  \beta_2 \\
	0&	-\beta_1  & -\beta_2  &  1\\
	\end{pmatrix}\begin{pmatrix}
        \phi_D' \\
		\phi_1' \\
		\phi_2'\\
		h'
	\end{pmatrix},
\end{equation}
where $\beta_{1,2}$ mix the SM Higgs with the singlet scalars, while $\alpha$ and $\theta_{1,2}$ parameterizes the mixing between dark sector fields. At first order in the scalar  angles the $\phi_D$ does not mix with the Higgs, and the predicted lepton EDM is given by an expression similar to Eq.~(\ref{eq:EDM}):
\begin{eqnarray}
		d_{l} &\simeq& \sum_{i=1,2} \frac{Y_{l} }{8 \pi^{2}v_i} \epsilon_{i}^{2} \tan\chi_{i}\ \beta_i c_{\theta}^{2}\, e\ \left(\frac{\log(x_{X_{i+1}\phi_i})}{x_{X_{i+1}\phi_i}-1}-\frac{\log(x_{X_{i+1}h})}{x_{X_{i+1}h}-1}\right).
\end{eqnarray}
A stable particle  $\chi$ in the dark sector contributes to the dark matter relic density, and its abundance can be thermally produced following the freeze-out of the annihilation process $\chi \chi\to {\rm SM}$ given by the tree-level exchange of a $X^{2,3}$ bosons that couple to the electromagnetic and the $Z$ currents. We are assuming that the mediator mass is larger than {twice the DM mass} $m_\chi$, so that the annihilation proceeds from an off-shell exchange {and the thermally averaged cross-section is of order
\begin{equation}\label{cross-section}
\langle \sigma_{\chi\chi \rightarrow {\rm SM}}v_{\rm rel}\rangle \sim \epsilon^{2} \alpha_{\rm EM} \alpha_{D} \frac{16 \pi m_{\chi}^{2}}{(4m_{\chi}^{2}-M_{X}^{2})^{2}},
\end{equation}
 where $\alpha_{D}=g_{D}^{2}/4 \pi$ and $v_{\rm rel}$ is the DM particles relative velocity.} The correct dark matter abundance is obtained when the freeze-out thermal cross-section is $\langle\sigma_{\chi\chi \rightarrow {\rm SM}}v_{\rm rel}\rangle \sim 1.7 \times 10^{-9} \;\mathrm{GeV}^{-2}$ \cite{bauer2018introductiondarkmatter}. Taking $m_\chi\lesssim 1$ GeV, $M_X\sim {\rm few}\times m_\chi$ and $g_D\sim \mathcal{O}(1)$, the observed relic abundance is given by $\epsilon\sim10^{-4}$, which as shown in Fig.~\ref{kinetic_mixing_plot}, is in the reach of the EDM searches.  However, indirect and direct detection experiments place strong constraints on this simple scenario. On one hand CMB anisotropies are sensitive to the energy injection in the intergalactic medium that could arise from DM annihilation, and for the thermal cross section needed to get the correct DM relic abundance, Planck data impose, for s-wave channel annihilation, an indirect lower limit on the DM mass  $m_\chi\gtrsim 30$ GeV \cite{Planck:2018vyg}. On the other hand, for DM masses above few GeVs, direct detection searches severely constrain the DM interactions with nuclei, which are parameterized by the kinetic mixing parameter $\epsilon$ \cite{LUXdarkmatter,XenonDirect,PandaX-4TDarkMatter,LZDarkMatter}. 
 
 The thermal freeze-out of a sub-GeV DM particle is possible in the so-called inelastic dark matter scenario \cite{Smith_2001}. The idea is that DM interacts with the SM only via an inelastic scattering between two states $\chi_{H,S}$ with a mass splitting $\delta m_\chi$. Hence, the annihilation channel $\chi_H \chi_S\to {\rm SM}$ sets the relic abundance when the process freezes out, but after the decay of the heavier state $\chi_H\to \chi_S +{\rm SM}$ only $\chi_S$ populates the Universe and it cannot annihilate with itself to give a signal in an indirect detection search. As a result, it is possible for dark matter candidates with masses at or below the GeV scale to elude these indirect bounds and in this energy region direct detection constraints are less severe (\cite{Abdelhameed_2019}, see however \cite{Aprile_2019}). In addition to this, bounds from direct detection experiments can be considerably relaxed by considering a mass splitting greater than $\delta m_{\chi} \sim  1 \div 100 \, \mathrm{keV}$, where the direct detection of DM rely on the effect of cosmic ray upscattering and so are much less severe \cite{Bell_2021}. Larger mass splitting are also cosmologically favoured by the requirement of sufficiently short-lived $\chi_{H}$, because for  $\delta_{\chi}<2m_e$ the only possible decay channel for the heavy state is $\chi_{H}\rightarrow \chi_{S} \nu \bar{\nu}$, easily resulting in a lifetime  greater than the age of the Universe \cite{Batell_2009}. 

 We can reproduce the inelastic scenario introducing two chiral $SU(2)_{D}$ doublets $\chi_L, \psi_R$, in addition to the field content presented at the beginning of this section. The relevant interactions that involve the extra fermions are the following
\begin{eqnarray}
    \mathcal{L} &\supset & -m_{D} \overline{\chi_{L}} \psi_{R} - \sum_{i=1,2} Y_{D,i} \overline{\chi_{L}} \Sigma_i \psi_{R}-\sum_{i=1,2} Y_{L,i}  \overline{\chi^{c}_{L}} i \sigma_{2} \Sigma_{i} \chi_{L} - \sum_{i=1,2} Y_{R,i}  \overline{\psi^{c}_{R}} i \sigma_{2} \Sigma_{i} \psi_{R} + {\rm h.c.} \nonumber \\
    &-& \frac{g_{D}}{2} \overline{\chi_{L}} \gamma_{\mu} \sigma^{a}X^{\mu}_{a} \chi_{L} - \frac{g_{D}}{2} \overline{\psi_{R}} \gamma_{\mu}\sigma^{a}X^{\mu}_{a} \psi_{R}.
\end{eqnarray}
The SU(2) components of the chiral doublets combine into two Dirac fermions $\Psi_H, \Psi_S$ (see Appendix \ref{app:iDM} for more details). For small values of the Yukawas $Y_i v_D\ll m_D$, the two fermions are approximately degenerate with a mass $m_\Psi \sim m_D$ and a small mass splitting $\delta_\Psi \sim v_D Y_i$. In the limit $Y_{L,i}=Y_{R,i}$ and negligible $Y_{D,i}$, the dark fermions exclusively have off-diagonal interactions with the $X_{2}$ gauge boson that, {together with $X_{3}$} , kinetically mixes with the SM hypercharge boson. The $X_{3}$ gauge interactions feature instead diagonal $\Psi_{H,S}$ currents which are unavoidable in the limit of small Yukawas. The elastic annihilations mediated by $X_3$ can be suppressed by taking $v_1\gg  v_2$ to enhance the $X_3$ mass (see Eq.~(\ref{eq:gbmasses})) and/or considering $\epsilon_1\ll \epsilon_2$, which is possible if in the UV the heavy states have small couplings with the scalar $\Sigma_1$. In the following, we assume this is the case and neglect the $X_3$-mediated elastic annihilations. Hence, the relic abundance and the EDM prediction will be dominated by the $X_2$ couplings, and we rename $\epsilon_2\equiv \epsilon$, $v_2\equiv v_D$, $M_{X_2}\equiv M_X$.

The relic density of $\Psi_H, \Psi_S$ following the freeze-out of the co-annihilations is determined by the thermal cross-section $\langle\sigma v\rangle $ for the scattering process $\Psi_H \Psi_S\to {\rm SM}$ and by the mass splitting $\delta_\Psi$. Since the $\Psi_H$ will eventually decay and annihilate into $\Psi_S$ plus SM particles, the relic abundance of $\Psi_S$ will be the sum of the  $\Psi_H$ and $\Psi_S$ densities at the time when the co-annihilations cease. In practice, to calculate the final DM abundance, we can define an effective thermal cross-section \cite{Griest:1990kh}
\begin{equation}
\langle \sigma_{\rm eff} v_{\rm rel}\rangle= \frac{\langle \sigma v_{\rm rel} \rangle}{2}\left(1-\frac{\delta_\Psi}{T}+\frac{3}{4}\frac{\delta_\Psi}{m_\Psi}+\mathcal{O}(\delta^2_\Psi)\right),\label{eq:effsigma}
\end{equation}
where we expanded the Boltzmann factors for small mass splittings\footnote{This is justified because the relic density is dominated by the equilibrium value at the freeze-out temperature} and where 
$\langle \sigma v_{\rm rel} \rangle$  is the total annihilation cross section into SM model particles, including leptonic and hadronic final states. 

The leptonic cross-section is similar to Eq.~(\ref{cross-section}) up to mass corrections, and to compute the hadronic cross section in the GeV and sub-GeV region, we considered $\langle \sigma_{\rm had} \rangle (s) = \langle \sigma_{\mu \mu} \rangle (s) R(s)$, where $R(s)$ is the experimental {ratio of $e^{+} e^{-}$ cross section to hadronic versus muon final states at colliders}
\cite{ParticleDataGroup:2020ssz}. Then, the correct DM abundance is obtained imposing $\langle \sigma_{\rm eff} v_{\rm rel}\rangle\sim  1.7\times 10^{-9}\ {\rm GeV^{-2}}$.  

The phenomenology of inelastic dark matter at laboratory experiments has been widely treated in literature and it is, in general, model dependent \cite{abdullahi2023semivisibledarkphotonphenomenology,Mohlabeng_2019,Izaguirre_2016, Izaguirre_2017}. Indeed, depending on the parameters of the model and the specific experiment, the signature of the dark sector at collider and beam dump experiments is different.
The decay rate of $\Psi_H$, in the limit $M_{X} > m_{\Psi}\gg m_{f}$, reads
\begin{equation}\label{decay}
    \sum_f\Gamma(\Psi_{H} \rightarrow \Psi_{S} f \bar{f}) = \sum_f\frac{{4}\epsilon^{2} \alpha \alpha_{D} \delta_{\Psi}^{5}}{15 \pi M_{X}^{4}}.
\end{equation}
Small mass splitting suppresses the decay rate of the heavy dark fermion. {In collider experiments, dark fermions are produced only through the coupling of the $X$ boson to the electromagnetic current \eq{eq:Xemcoupl}.} If the lifetime of $\Psi_{H}$ is short enough, its signature will be characterized by a pair of fermions plus missing energy, given by the decay chain $X \rightarrow \Psi_{H} \Psi_{S} \rightarrow \Psi_{S}\Psi_{S} f \bar{f}$. On the other hand, if $\Psi_{H}$ is sufficiently long-lived, it will decay outside the detector rendering the final state completely invisible. The latter scenario is possible if $\delta_\Psi\lesssim 0.01\times m_\Psi$ with $m_\Psi$ in the sub-GeV region, and the relevant experimental bounds on the model parameters arise from searches for a dark boson decaying to an invisible final state \cite{BaBar_invisible_Lees_2017, BESIII_Ablikim_2023, NA64_invisible_Andreev_2021, NA64:2016oww, NA62:2019meo}. We focus on the small mass splitting case because it is {less constrained by present experiments}, while for the semi-visible signatures beam-dump experiments can place strong constraints on the model in the case of sub-GeV DM mass \cite{Izaguirre_2017}. In our case, even for a long-lived $\Psi_H$, this region of the parameter space is testable by the upcoming searches of the electron EDM.

We perform a scan over the parameter space of the model, varying the parameters as described in Table \ref{tab:scan}. We fix the mass splitting to be $\delta_\psi\lesssim 0.01\times m_\psi$, but require $\delta_\Psi > 2m_e$ to avoid the BBN bounds on the $\Psi_H$ lifetime. Furthermore, we consider the experimental constraints on the kinetic mixing parameter $\epsilon$ from the dark boson to invisible decays. We also assume and enforce that $M_{X}> 2 m_{\Psi}$, so that the DM  abundance is set by the $\Psi_{H,S}$ co-annihilation into the SM rather than from the gauge interactions within the dark sector, which makes the model more predictive.

\begin{table}[ht]
	\begin{center}
		\begin{tabular}{c c c }
			\toprule
			Parameter & Lower limit & Upper limit\\
			\midrule
			\qquad$\beta_{1,2}$ &\qquad $10^{-4}$& \qquad$10^{-2}$ \\
			\qquad$\epsilon$ &\qquad $10^{-6}$ & \qquad\cite{BaBar_invisible_Lees_2017, BESIII_Ablikim_2023, NA64_invisible_Andreev_2021, NA64:2016oww, NA62:2019meo}\\
			\qquad$g_{D}$ &\qquad $10^{-2}$ &\qquad 1\\
            \qquad$\tan {\chi}$ &\qquad $10^{-2}$ &\qquad 1\\
			\qquad$v_D$ & \qquad 1 GeV & \qquad20 GeV\\
            \qquad$m_{\phi}$ & \qquad 0.1 GeV & \qquad 50 GeV\\
            \bottomrule
	\end{tabular}	\end{center} 
	\caption{Table representing the range of the scanned parameter values. Initial and final values are chosen in accordance of the preliminary analysis for a sizeable eEDM and current constraints from laboratory experiments.}
		\label{tab:scan}
\end{table}

 \begin{figure}[t]
	\centering
	\includegraphics[width = 0.8\linewidth]{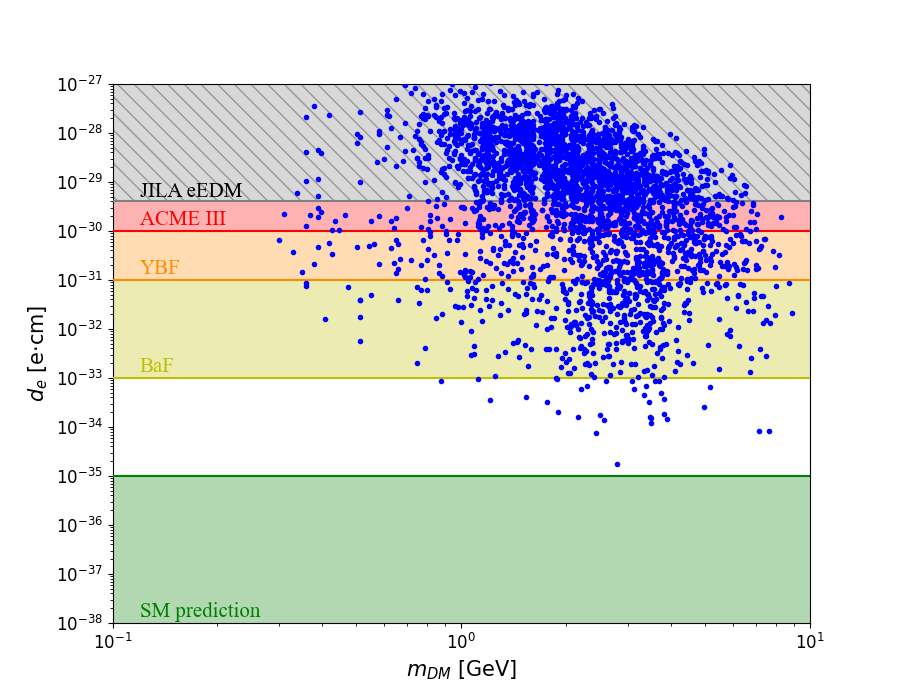}
	\caption{Scatter plot representing on the $y$-axis the eEDM and on the $x$-axis the DM mass satisfying the correct relic abundance. Blue points represent the predicted eEDM and DM mass for parameters randomly chosen accordingly to Tab.\ref{tab:scan}. Gray region represents the already excluded parameter space by JILA eEDM collaboration. Colored regions between $10^{-33}$ and 4.1 $\times 10^{-30}$ show the future eEDM sensitivities. Finally, green region covers the SM prediction for the equivalent eEDM \cite{SMpredictioneEDM}.}
	\label{scatter_plot}
\end{figure}
 In Fig.~\ref{scatter_plot} we show the predicted eEDM and DM mass for each of the parameter space points. The DM mass is obtained by fixing the effective cross-section of Eq.~(\ref{eq:effsigma}) to the value $\sim 1.7\times 10^{-9}$ ${\rm GeV}^{-2}$ of the thermal cross-section needed to have the correct relic abundance. We show in Fig.~\ref{scatter_plot} that, in a large region of the parameter space, the predicted eEDM lies between the current experimental bounds and the future experimental sensitivity, making the model testable in future eEDM searches.

\section{Summary}\label{sec:concl}
Non-Abelian dark sectors can have a naturally small portal to the visible sector and simultaneously offer pathways to address unresolved puzzles such as the observed dark matter abundance and baryon asymmetry in the Universe.
In this work, we have explored the phenomenology of a CP-odd portal (defined in Eq.~(\ref{eq:nonabelianmixingops})) to a non-Abelian dark sector. We have argued that in the presence of large CP-violating phases in the UV, the CP-odd portal is closely related to the kinetic mixing between the dark and the SM gauge bosons. We have focused on the contribution to electric dipole moments, as they are the most sensitive observables to CP violation. Hence, we have discussed the potential of current and future EDM searches in probing this scenario.

In Section \ref{sec:nonabelianmix}, we have discussed the features of the vector mixing portals under general assumptions and determined the sensitivity of electron EDM searches to the CP-odd operator coefficient, specifically in the case of a single massive dark gauge boson mixing with the hypercharge vector. Assuming comparable coefficients for the CP-even and CP-odd operators of Eq.~(\ref{eq:nonabelianmixingops}), we have compared the EDM sensitivity with the reach of dark photon searches for the kinetic mixing parameter $\epsilon$ as a function of the dark photon mass. We have identified regions of parameter space, particularly for $M_X \gtrsim 1$ GeV, where future EDM searches are competitive with, or even more sensitive than, other experimental probes (see Fig. \ref{kinetic_mixing_plot}).

In Section \ref{sec:DSmodels}, we have specialized the discussion to a dark gauge group $SU(2)_D$ with specific scalar contents. In subsection \ref{model1}, we have considered a single triplet scalar that acquires a vacuum expectation value, and an Abelian subgroup $U(1)_D$ survives the spontaneous symmetry breaking. As a result of the gauge boson kinetic mixing $\epsilon$, particles charged under the extra $U(1)_D $ acquire an $\epsilon$-dependent millicharge. Since $\epsilon$ parameterizes the prediction for the electron electric dipole moment, we have found an indirect sensitivity of EDM experiments to the millicharge. Once again, we have identified regions where the EDM reach provides interesting constraints compared to other experimental searches for millicharged particles (see Fig. \ref{millicharge_plot}).

In Subsection \ref{model2}, we have augmented the scalar sector to completely break the $ SU(2)_D $ group, resulting in {three} massive dark gauge bosons. We have implemented an inelastic dark matter scenario where the DM abundance is closely related to the size of $\epsilon$, among other model parameters. We have explored whether future eEDM searches, considering all experimental constraints, could potentially exclude this scenario. As shown in Fig.~\ref{scatter_plot}, we have found that for many points in the parameter space, compatible with the observed DM abundance and various experimental bounds, a potentially observable eEDM is predicted.

\section*{Acknowledgments}
This work is supported by the Generalitat Valenciana project CIPROM/2021/054 and CIPROM/2022/66 and the Spanish AEI-MICINN PID2020-113334GB-I00 \\
(AEI/10.13039/501100011033).
\clearpage

\appendix
\section{Non-Abelian kinetic mixing from the UV}\label{app:UV}
\begin{figure}[t]
	\centering
	\includegraphics[width = 0.5\linewidth]{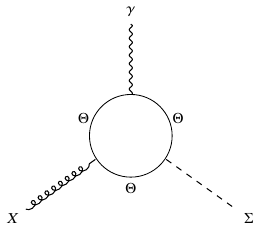}
	\caption{One loop diagram generating the CP-even and CP-odd terms in Eq.~(\ref{eq:nonabelianmixingops}).}
	\label{UV_completion}
\end{figure}

The effective terms in Eq.~\eqref{eq:nonabelianmixingops} require an anomaly free UV completion. The simplest choice consists in introducing vector-like heavy chiral fermions $\Theta$ in the defining representation of $G_{D}$\footnote{ The generators $T^a$ of the fundamental representation are normalised such that $\Tr(T^a T^b)=\delta^{ab}/2$} and having hypercharge $Y$. The relevant Lagrangian terms are the following
\begin{gather}
	\mathcal{L}_{\Theta} \supset \
	-g_1 Y \bar{\Theta}_i\gamma_{\mu} \Theta_i B^{\mu} - g_{D} \bar{\Theta}_i T^{a} \gamma_{\mu} \Theta_i X^{\mu}_{a} -M_{ii} \bar{\Theta}_{iR} \Theta_{iL} - \mathcal{Y}_{ij} \bar{\Theta}_{iR} \Sigma^{a} T^{a} \Theta_{jL} + {\rm h.c.},\label{eq:UVlag}
\end{gather}
where $g_1=e/c_\theta$ is the hypercharge coupling, and $\mathcal{Y}_{ij}$ is a complex Yukawa-like matrix, which {in the presence of a diagonal real mass, $M_{ii}$, has physical phases \cite{Botella:2004ks,Davidson:2005cw}}. The effective operators in Eq.~(\ref{eq:nonabelianmixingops}) of the text, originate from the one loop diagram shown in Fig.~\ref{UV_completion}, after  the heavy degrees of freedom have been integrated out. 
Matching the diagram of Fig.~\ref{UV_completion} for a single heavy fermion having a sizable (complex) Yukawa coupling $\mathcal{Y}$ and mass $M$, gives the following coefficient for the operators in Eq.~(\ref{eq:nonabelianmixingops})
\begin{align}\label{eq.A.2}
	&\frac{C}{\Lambda} = \frac{g_{D}g_1 Y \mathrm{Re}[\mathcal{Y}]}{12 \pi^{2} M}; \quad \quad 
	\frac{\tilde{C}}{\Lambda} = \frac{g_{D}g_1 Y \mathrm{Im}[\mathcal{Y}]}{16 \pi^{2} M},
\end{align}
For the one heavy fermion case, we find that 
\begin{equation}\label{UVrelation}
	\tilde{\epsilon} = \frac{3}{4} \frac{\mathrm{Im}[\mathcal{Y}]}{\mathrm{Re}[\mathcal{Y}]} \epsilon \equiv (\tan \chi)\, \epsilon,
\end{equation}
having defined $\chi$ as the phase of $\mathcal{Y}$ up to an $\sim \mathcal{O}(1)$ factor. In general, the value of $\tan \chi$ is UV-dependent, so in the text we just assume that $\tan \chi\sim \mathcal{O}(1)$. 

Since we assume to have fermion with hypercharges, collider and cosmological bounds may apply. To obtain kinetic mixings $\epsilon \sim 10^{-3} \div 10^{-5}$, we need $M\gtrsim 1$ TeV, which could be close to the reach of LHC searches. A detailed analysis is beyond the scope of this work.
In addition, to avoid the strict cosmological bounds on the relic abundance of charged particles, we need the heavy fermions to decay into standard model and dark sector particles. For instance, in the case studied in Section \ref{model2} where the DM candidate is the fundamental representation of $SU(2)_D$, it is sufficient to consider $\Theta$ transforming in the $(\underline{2}, 1/2, \underline{2})$ representation of $SU(2)_L\times U(1)_Y\times SU(2)_D$. Then, the $\Theta$ can decay into the DM$+$SM via the portal interaction
\begin{equation}
    \lambda \bar{\Theta} H \chi+{\rm h.c.}
\end{equation}

\section{Matching contributions to the EDM in the UV} \label{app:barr-zee}
\begin{figure}[t]
	\centering
	\includegraphics[width = 0.5\linewidth]{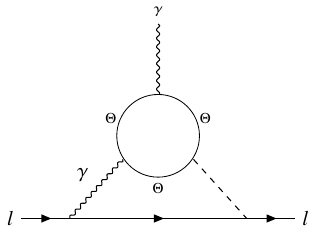}
	\caption{Barr-Zee diagram arising from the UV completion of the effective theory.}
	\label{Barr_zee}
\end{figure}

By adding heavy chiral fermions with hypercharge, there could also be a direct matching contribution to the dipole operators arising from the two-loop Barr-Zee diagrams of Fig.~\ref{Barr_zee}.

We neglect the diagram with a $Z$ exchange because, like in the diagram of Fig.~\ref{fig_feynman_EDM}, it is suppressed by the small $Z$ lepton vector coupling. The heavy fermions have the interactions described in Eq.~(\ref{eq:UVlag}), and we once again consider the case where only one has a sizable complex Yukawa coupling $\mathcal{Y}$ with the adjoint scalar $\Sigma$.  Taking for simplicity one dark scalar $\phi^a$, mixing with the Higgs with a small angle $\beta^a$, the Barr-Zee contribution to the electric dipole moment is the following \cite{Chang:1993kw}
\begin{equation}
    d_{l}= \frac{q^2_\Theta e\alpha}{8\pi^3}\frac{Y_l\Im[\mathcal{Y}]}{M}\beta^a\left[g\left(\frac{M^2}{m^2_{\phi^a}}\right)-g\left(\frac{M^2}{m^2_{h}}\right)\right], \label{eq:BarrZee}
\end{equation}
where $q_\Theta$ is the $\Theta$ electric charge,  $Y_l$ is the lepton SM Yukawa and $g(z)$ is defined as
\begin{equation}
    g(z)=\frac{z}{2}\int_0^1 dx\frac{1}{x(1-x)-z}\log\left(\frac{x(1-x)}{z}\right).
\end{equation}
In the limit $z\gg 1$, we can approximate the integrand as $-1/2\log(x(1-x)/z)$, so that Eq.~(\ref{eq:BarrZee}) becomes:
\begin{equation}
    d_{l}\sim \frac{q^2_\Theta e\alpha}{16\pi^3}\frac{Y_l\Im[\mathcal{Y}]}{M}\beta^a\log\left(\frac{m^2_h}{m^2_{\phi^a}}\right).
\end{equation}
The resulting EDM has a different parametric dependence with respect to the contribution of the CP-odd effective operator. In particular, it does not depend on the dark gauge coupling $g_D$. Therefore, in the UV-complete theory, it is possible that the CP-odd portal contribution is suppressed, for instance by taking $g_D \ll 1$, but the contribution to the EDM could still be sizable from Barr-Zee like diagrams. 

\section{Fermion masses in the $SU(2)_D$ model\label{app:iDM}}
In this section we give more details about the fermion mass diagonalization for the iDM model of Section \ref{model2}. Introducing two Weyl fermion doublets with opposite chirality
\begin{equation}
		\chi_{L} = \begin{pmatrix}
		 \chi_{L,1} \\
		 \chi_{L,2} 
	\end{pmatrix}; \quad \psi_{R}=\begin{pmatrix}
		 \psi_{R,1} \\
		 \psi_{R,2}
	\end{pmatrix},
\end{equation}
the dark sector interactions relevant for these extra fermions read
\begin{eqnarray}
    \mathcal{L} &\supset & -m_{D} \overline{\chi_{L}} \psi_{R} - \sum_{i=1,2} Y_{D,i} \overline{\chi_{L}} \Sigma_i \psi_{R}-\sum_{i=1,2} Y_{L,i}  \overline{\chi^{c}_{L}} i \sigma_{2} \Sigma_{i} \chi_{L} - \sum_{i=1,2} Y_{R,i}  \overline{\psi^{c}_{R}} i \sigma_{2} \Sigma_{i} \psi_{R} + {\rm h.c.} \nonumber \\
    &-& \frac{g_{D}}{2} \overline{\chi_{L}} \gamma_{\mu} \sigma^{a}X^{\mu}_{a} \chi_{L} - \frac{g_{D}}{2} \overline{\psi_{R}} \gamma_{\mu}\sigma^{a}X^{\mu}_{a} \psi_{R}.
\end{eqnarray}

Upon SSB, when the scalars acquire vacuum expectation values as specified in Eq.~\eqref{vacuumexpvalue}, the fermions receive additional mass terms from the Yukawa couplings. Taking the right-handed and left-handed Yukawas to be the same ($Y_{L,1}=Y_{R,1}=Y_{1}$, $Y_{L,2}=Y_{R,2}=Y_{2}$) and neglecting the $Y_{D,i}$ contribution, which is the relevant limit for the inelastic scenario,
we have the following mass matrix for the fermions
\begin{equation}
	\begin{pmatrix}
		 \overline{\chi_{L,1}} & \overline{\chi_{L,2}}& \overline{\psi^{c}_{R,1}}& \overline{\psi^{c}_{R,2}}
	\end{pmatrix}\begin{pmatrix}
		i Y_{1} v_1 & -Y_{2} v_2 & m_{D} & 0 \\
		-Y_{2} v_2 & i Y_{1} v_1 &  0 & m_{D}\\
		m_{D} & 0  & i Y_{1} v_1& -Y_{2} v_2 \\
            0  & m_{D}  & -Y_{2} v_2& i Y_{1} v_1\\
	\end{pmatrix}\begin{pmatrix}
		\chi^{c}_{L,1} \\
		\chi^{c}_{L,2}\\
		\psi_{R,1}\\
            \psi_{R,2}
	\end{pmatrix}.
\end{equation}
Since $Y_{1}$ does not affect neither the mass splitting nor the structure of the final gauge interactions, we consider for simplicity $Y_{1} v_1 \ll Y_{2} v_2$, and the eigenstates and mass spectrum  thus read
 \begin{eqnarray}
     \chi_{R,H_{1}} &=& \frac{1}{2}\left( - i \chi^{c}_{L,1} -i \chi^{c}_{L,2} + i\psi_{R,1} + i \psi_{R,2}\right), \nonumber\\ 
    \chi_{R,H_{2}} &=& \frac{1}{2}\left( -\chi^{c}_{L,1} + \chi^{c}_{L,2} -\psi_{R,1} +\psi_{R,2}\right), \nonumber\\ \nonumber\\
     \chi_{R,S_{1}} &=& \frac{1}{2}\left( i \chi^{c}_{L,1} - i\chi^{c}_{L,2} - i \psi_{R,1} + i \psi_{R,2}\right), \nonumber\\
     \chi_{R,S_{2}} &=& \frac{1}{2}\left(   \chi^{c}_{L,1} +  \chi^{c}_{L,2} + \psi_{R,1} + \psi_{R,2}\right), \nonumber\\  \nonumber\\
     m_{S_{1,2}} = m_{D} &-& v_2 Y_{2} = M_{1}  ;\quad m_{H_{1,2}} = m_{D} + v_2 Y_{2} = M_{2},
 \end{eqnarray}
resulting in a mass splitting $\delta_{\chi} = 2v_2Y_{2}$. The pairs of mass degenerate Majorana states combine in two Dirac fermions defined as

\begin{equation}
    \Psi_{H} = \begin{pmatrix}
        \chi^{c}_{H} \\
        \xi_{H}
    \end{pmatrix};\quad \Psi_{S} = \begin{pmatrix}
        \xi^{c}_{S} \\
        \chi_{S}
    \end{pmatrix}; \quad \chi_{H,S} = \frac{\chi^{1}_{H,S} + i \chi^{2}_{H,S}}{\sqrt{2}};
    \quad \xi_{H,S} = \frac{\chi^{1}_{H,S} - i \chi^{2}_{H,S}}{\sqrt{2}}\label{eq:massmatrix}
\end{equation} 
with masses
\begin{equation}
    \mathcal{L}_{M} = -M_{1} \overline{\Psi}_{H} \Psi_{H} - M_{2} \overline{\Psi}_{S} \Psi_{S}.
\end{equation}
The gauge interactions in the $\Psi_{H,S}$ basis are the following:

\begin{align}
    &\frac{g_D}{2}\Big( \overline{\Psi}_{H} \gamma_{\mu}\Psi_{S}(X^{\mu}_{1} + i X_{2}^{\mu}) + \frac{1}{2}(\overline{\Psi}_{H} \gamma_{\mu} \Psi_{H}  - \overline{\Psi}_{S} \gamma_{\mu} \Psi_{S})X_{3}^{\mu}  + {\rm h.c.} \Big),\nonumber
\end{align}
where the $X_{1}^{\mu},X_{2}^{\mu}$  gauge boson couple to the fermions only through an off-diagonal vector coupling, while $X_{3}^{\mu}$ presents only diagonal currents. Diagonal currents would also be  present in the $X_2$ couplings for $Y_{L,i}-Y_{R,i}\neq0$ and $Y_{D,i}\neq 0$, which motivates the limit we consider in Eq.~(\ref{eq:massmatrix}) to avoid the elastic annihilations. 

\bibliography{references}

\providecommand{\href}[2]{#2}\begingroup\raggedright\begin{thebibliography}{100}

\bibitem{Pospelov:2007mp}
M.~Pospelov, A.~Ritz and M.B.~Voloshin, \emph{{Secluded WIMP Dark Matter}},
  \href{https://doi.org/10.1016/j.physletb.2008.02.052}{\emph{Phys. Lett. B}
  {\bfseries 662} (2008) 53} [\href{https://arxiv.org/abs/0711.4866}{{\ttfamily
  0711.4866}}].

\bibitem{Arkani-Hamed:2008hhe}
N.~Arkani-Hamed, D.P.~Finkbeiner, T.R.~Slatyer and N.~Weiner, \emph{{A Theory
  of Dark Matter}},
  \href{https://doi.org/10.1103/PhysRevD.79.015014}{\emph{Phys. Rev. D}
  {\bfseries 79} (2009) 015014}
  [\href{https://arxiv.org/abs/0810.0713}{{\ttfamily 0810.0713}}].

\bibitem{Haba:2010bm}
N.~Haba and S.~Matsumoto, \emph{{Baryogenesis from Dark Sector}},
  \href{https://doi.org/10.1143/PTP.125.1311}{\emph{Prog. Theor. Phys.}
  {\bfseries 125} (2011) 1311}
  [\href{https://arxiv.org/abs/1008.2487}{{\ttfamily 1008.2487}}].

\bibitem{Shelton:2010ta}
J.~Shelton and K.M.~Zurek, \emph{{Darkogenesis: A baryon asymmetry from the
  dark matter sector}},
  \href{https://doi.org/10.1103/PhysRevD.82.123512}{\emph{Phys. Rev. D}
  {\bfseries 82} (2010) 123512}
  [\href{https://arxiv.org/abs/1008.1997}{{\ttfamily 1008.1997}}].

\bibitem{Chu:2011be}
X.~Chu, T.~Hambye and M.H.G.~Tytgat, \emph{{The Four Basic Ways of Creating
  Dark Matter Through a Portal}},
  \href{https://doi.org/10.1088/1475-7516/2012/05/034}{\emph{JCAP} {\bfseries
  05} (2012) 034} [\href{https://arxiv.org/abs/1112.0493}{{\ttfamily
  1112.0493}}].

\bibitem{Buttazzo:2019iwr}
D.~Buttazzo, L.~Di~Luzio, G.~Landini, A.~Strumia and D.~Teresi, \emph{{Dark
  Matter from self-dual gauge/Higgs dynamics}},
  \href{https://doi.org/10.1007/JHEP10(2019)067}{\emph{JHEP} {\bfseries 10}
  (2019) 067} [\href{https://arxiv.org/abs/1907.11228}{{\ttfamily
  1907.11228}}].

\bibitem{Buttazzo:2019mvl}
D.~Buttazzo, L.~Di~Luzio, P.~Ghorbani, C.~Gross, G.~Landini, A.~Strumia et~al.,
  \emph{{Scalar gauge dynamics and Dark Matter}},
  \href{https://doi.org/10.1007/JHEP01(2020)130}{\emph{JHEP} {\bfseries 01}
  (2020) 130} [\href{https://arxiv.org/abs/1911.04502}{{\ttfamily
  1911.04502}}].

\bibitem{Landini:2020daq}
G.~Landini and J.-W.~Wang, \emph{{Dark Matter in scalar Sp($ \mathcal{N} $)
  gauge dynamics}}, \href{https://doi.org/10.1007/JHEP06(2020)167}{\emph{JHEP}
  {\bfseries 06} (2020) 167}
  [\href{https://arxiv.org/abs/2004.03299}{{\ttfamily 2004.03299}}].

\bibitem{Ardu:2020qmo}
M.~Ardu, L.~Di~Luzio, G.~Landini, A.~Strumia, D.~Teresi and J.-W.~Wang,
  \emph{{Axion quality from the (anti)symmetric of SU($ \mathcal{N} $)}},
  \href{https://doi.org/10.1007/JHEP11(2020)090}{\emph{JHEP} {\bfseries 11}
  (2020) 090} [\href{https://arxiv.org/abs/2007.12663}{{\ttfamily
  2007.12663}}].

\bibitem{Frigerio:2022kyu}
M.~Frigerio, N.~Grimbaum-Yamamoto and T.~Hambye, \emph{{Dark matter from the
  centre of SU(N)}},
  \href{https://doi.org/10.21468/SciPostPhys.15.4.177}{\emph{SciPost Phys.}
  {\bfseries 15} (2023) 177}
  [\href{https://arxiv.org/abs/2212.11918}{{\ttfamily 2212.11918}}].

\bibitem{Borah:2022phw}
D.~Borah, E.~Ma and D.~Nanda, \emph{{Dark SU(2) gauge symmetry and scotogenic
  Dirac neutrinos}},
  \href{https://doi.org/10.1016/j.physletb.2022.137539}{\emph{Phys. Lett. B}
  {\bfseries 835} (2022) 137539}
  [\href{https://arxiv.org/abs/2204.13205}{{\ttfamily 2204.13205}}].

\bibitem{Soni:2016gzf}
A.~Soni and Y.~Zhang, \emph{{Hidden SU(N) Glueball Dark Matter}},
  \href{https://doi.org/10.1103/PhysRevD.93.115025}{\emph{Phys. Rev. D}
  {\bfseries 93} (2016) 115025}
  [\href{https://arxiv.org/abs/1602.00714}{{\ttfamily 1602.00714}}].

\bibitem{Yamanaka:2019yek}
N.~Yamanaka, H.~Iida, A.~Nakamura and M.~Wakayama, \emph{{Glueball scattering
  cross section in lattice SU(2) Yang-Mills theory}},
  \href{https://doi.org/10.1103/PhysRevD.102.054507}{\emph{Phys. Rev. D}
  {\bfseries 102} (2020) 054507}
  [\href{https://arxiv.org/abs/1910.07756}{{\ttfamily 1910.07756}}].

\bibitem{Holdom:1985ag}
B.~Holdom, \emph{{Two U(1)'s and Epsilon Charge Shifts}},
  \href{https://doi.org/10.1016/0370-2693(86)91377-8}{\emph{Phys. Lett. B}
  {\bfseries 166} (1986) 196}.

\bibitem{KinMix0}
F.~Chen, J.M.~Cline and A.R.~Frey, \emph{Non-abelian dark matter: Models and
  constraints},
  \href{https://doi.org/10.1103/physrevd.80.083516}{\emph{Physical Review D}
  {\bfseries 80} (2009) }.

\bibitem{KinMix1}
C.-W.~Chiang, T.~Nomura and J.~Tandean, \emph{{Nonabelian Dark Matter with
  Resonant Annihilation}},
  \href{https://doi.org/10.1007/JHEP01(2014)183}{\emph{JHEP} {\bfseries 01}
  (2014) 183} [\href{https://arxiv.org/abs/1306.0882}{{\ttfamily 1306.0882}}].

\bibitem{KinMix2}
J.M.~Cline and A.R.~Frey, \emph{{Nonabelian dark matter models for 3.5 keV
  X-rays}}, \href{https://doi.org/10.1088/1475-7516/2014/10/013}{\emph{JCAP}
  {\bfseries 10} (2014) 013} [\href{https://arxiv.org/abs/1408.0233}{{\ttfamily
  1408.0233}}].

\bibitem{KinMix3}
K.~Cheung, W.-C.~Huang and Y.-L.S.~Tsai, \emph{{Non-abelian Dark Matter
  Solutions for Galactic Gamma-ray Excess and Perseus 3.5 keV X-ray Line}},
  \href{https://doi.org/10.1088/1475-7516/2015/05/053}{\emph{JCAP} {\bfseries
  05} (2015) 053} [\href{https://arxiv.org/abs/1411.2619}{{\ttfamily
  1411.2619}}].

\bibitem{KinMix4}
G.~Barello, S.~Chang and C.A.~Newby, \emph{{Correlated signals at the energy
  and intensity frontiers from non-Abelian kinetic mixing}},
  \href{https://doi.org/10.1103/PhysRevD.94.055018}{\emph{Phys. Rev. D}
  {\bfseries 94} (2016) 055018}
  [\href{https://arxiv.org/abs/1511.02865}{{\ttfamily 1511.02865}}].

\bibitem{KinMix5}
J.~Choquette and J.M.~Cline, \emph{{Minimal non-Abelian model of atomic dark
  matter}}, \href{https://doi.org/10.1103/PhysRevD.92.115011}{\emph{Phys. Rev.
  D} {\bfseries 92} (2015) 115011}
  [\href{https://arxiv.org/abs/1509.05764}{{\ttfamily 1509.05764}}].

\bibitem{KinMix6}
C.D.~Carone, S.~Chaurasia and T.V.B.~Claringbold, \emph{{Dark sector portal
  with vectorlike leptons and flavor sequestering}},
  \href{https://doi.org/10.1103/PhysRevD.99.015009}{\emph{Phys. Rev. D}
  {\bfseries 99} (2019) 015009}
  [\href{https://arxiv.org/abs/1807.05288}{{\ttfamily 1807.05288}}].

\bibitem{KinMix7}
F.~Elahi and M.~Mohammadi~Najafabadi, \emph{{Neutron Decay to a Non-Abelian
  Dark Sector}}, \href{https://doi.org/10.1103/PhysRevD.102.035011}{\emph{Phys.
  Rev. D} {\bfseries 102} (2020) 035011}
  [\href{https://arxiv.org/abs/2005.00714}{{\ttfamily 2005.00714}}].

\bibitem{KinMix8}
P.~Ko, T.~Nomura and H.~Okada, \emph{{Dark matter physics in dark $SU(2)$ gauge
  symmetry with non-Abelian kinetic mixing}},
  \href{https://doi.org/10.1103/PhysRevD.103.095011}{\emph{Phys. Rev. D}
  {\bfseries 103} (2021) 095011}
  [\href{https://arxiv.org/abs/2007.08153}{{\ttfamily 2007.08153}}].

\bibitem{KinMix9}
T.~Nomura and H.~Okada, \emph{{Radiative neutrino mass model in dark
  non-Abelian gauge symmetry}},
  \href{https://doi.org/10.1103/PhysRevD.105.075010}{\emph{Phys. Rev. D}
  {\bfseries 105} (2022) 075010}
  [\href{https://arxiv.org/abs/2106.10451}{{\ttfamily 2106.10451}}].

\bibitem{KinMix10}
T.G.~Rizzo, \emph{{Kinetic mixing, dark Higgs triplets, and MW}},
  \href{https://doi.org/10.1103/PhysRevD.106.035024}{\emph{Phys. Rev. D}
  {\bfseries 106} (2022) 035024}
  [\href{https://arxiv.org/abs/2206.09814}{{\ttfamily 2206.09814}}].

\bibitem{KinMix11}
H.~Zhou, \emph{{Quasi-degenerate dark photon and dark matter}},
  \href{https://doi.org/10.1016/j.nuclphysb.2024.116474}{\emph{Nucl. Phys. B}
  {\bfseries 1000} (2024) 116474}
  [\href{https://arxiv.org/abs/2209.08843}{{\ttfamily 2209.08843}}].

\bibitem{KinMix12}
Y.~Cheng, X.-G.~He, F.~Huang, J.~Sun and Z.-P.~Xing, \emph{{Dark photon kinetic
  mixing effects for the CDF W-mass measurement}},
  \href{https://doi.org/10.1103/PhysRevD.106.055011}{\emph{Phys. Rev. D}
  {\bfseries 106} (2022) 055011}
  [\href{https://arxiv.org/abs/2204.10156}{{\ttfamily 2204.10156}}].

\bibitem{Alonso-Alvarez:2023rjq}
G.~Alonso-\'Alvarez, R.~Cao, J.M.~Cline, K.~Moorthy and T.~Xiao,
  \emph{{Nonabelian Kinetic Mixing in a Confining Phase}},
  \href{https://arxiv.org/abs/2309.13105}{{\ttfamily 2309.13105}}.

\bibitem{Fuyuto:2019vfe}
K.~Fuyuto, X.-G.~He, G.~Li and M.~Ramsey-Musolf, \emph{{CP-violating Dark
  Photon Interaction}},
  \href{https://doi.org/10.1103/PhysRevD.101.075016}{\emph{Phys. Rev. D}
  {\bfseries 101} (2020) 075016}
  [\href{https://arxiv.org/abs/1902.10340}{{\ttfamily 1902.10340}}].

\bibitem{Cheng:2021qbl}
Y.~Cheng, X.-G.~He, M.J.~Ramsey-Musolf and J.~Sun, \emph{{CP-violating dark
  photon kinetic mixing and type-III seesaw model}},
  \href{https://doi.org/10.1103/PhysRevD.105.095010}{\emph{Phys. Rev. D}
  {\bfseries 105} (2022) 095010}
  [\href{https://arxiv.org/abs/2104.11563}{{\ttfamily 2104.11563}}].

\bibitem{Roussy:2022cmp}
T.S.~Roussy et~al., \emph{{An improved bound on the electron\textquoteright{}s
  electric dipole moment}},
  \href{https://doi.org/10.1126/science.adg4084}{\emph{Science} {\bfseries 381}
  (2023) adg4084} [\href{https://arxiv.org/abs/2212.11841}{{\ttfamily
  2212.11841}}].

\bibitem{ACMEIII_Hiramoto_2023}
A.~Hiramoto, T.~Masuda, D.~Ang, C.~Meisenhelder, C.~Panda, N.~Sasao et~al.,
  \emph{Sipm module for the acme iii electron edm search},
  \href{https://doi.org/10.1016/j.nima.2022.167513}{\emph{Nuclear Instruments
  and Methods in Physics Research Section A: Accelerators, Spectrometers,
  Detectors and Associated Equipment} {\bfseries 1045} (2023) 167513}.

\bibitem{YBFFitch_2021}
N.J.~Fitch, J.~Lim, E.A.~Hinds, B.E.~Sauer and M.R.~Tarbutt, \emph{Methods for
  measuring the electron’s electric dipole moment using ultracold ybf
  molecules}, \href{https://doi.org/10.1088/2058-9565/abc931}{\emph{Quantum
  Science and Technology} {\bfseries 6} (2020) 014006}.

\bibitem{BaFPhysRevA.98.032513}
A.C.~Vutha, M.~Horbatsch and E.A.~Hessels, \emph{Orientation-dependent
  hyperfine structure of polar molecules in a rare-gas matrix: A scheme for
  measuring the electron electric dipole moment},
  \href{https://doi.org/10.1103/PhysRevA.98.032513}{\emph{Phys. Rev. A}
  {\bfseries 98} (2018) 032513}.

\bibitem{Hook:2010tw}
A.~Hook, E.~Izaguirre and J.G.~Wacker, \emph{{Model Independent Bounds on
  Kinetic Mixing}}, \href{https://doi.org/10.1155/2011/859762}{\emph{Adv. High
  Energy Phys.} {\bfseries 2011} (2011) 859762}
  [\href{https://arxiv.org/abs/1006.0973}{{\ttfamily 1006.0973}}].

\bibitem{ATLAS:2023jyp}
{\scshape ATLAS} collaboration, \emph{{Search for dark photons in rare $Z$
  boson decays with the ATLAS detector}},
  \href{https://doi.org/10.1103/PhysRevLett.131.251801}{\emph{Phys. Rev. Lett.}
  {\bfseries 131} (2023) 251801}
  [\href{https://arxiv.org/abs/2306.07413}{{\ttfamily 2306.07413}}].

\bibitem{Ferber_2024}
T.~Ferber, A.~Grohsjean and F.~Kahlhoefer, \emph{Dark higgs bosons at
  colliders}, \href{https://doi.org/10.1016/j.ppnp.2024.104105}{\emph{Progress
  in Particle and Nuclear Physics} {\bfseries 136} (2024) 104105}.

\bibitem{ATLAS2022}
G.~Aad, B.~Abbott, D.C.~Abbott, K.~Abeling, S.H.~Abidi, A.~Aboulhorma et~al.,
  \emph{A detailed map of higgs boson interactions by the atlas experiment ten
  years after the discovery},
  \href{https://doi.org/10.1038/s41586-022-04893-w}{\emph{Nature} {\bfseries
  607} (2022) 52–59}.

\bibitem{CMS:2022dwd}
{\scshape CMS} collaboration, \emph{{A portrait of the Higgs boson by the CMS
  experiment ten years after the discovery.}},
  \href{https://doi.org/10.1038/s41586-022-04892-x}{\emph{Nature} {\bfseries
  607} (2022) 60} [\href{https://arxiv.org/abs/2207.00043}{{\ttfamily
  2207.00043}}].

\bibitem{Atlas22022}
G.~Aad, B.~Abbott, D.C.~Abbott, A.~Abed~Abud, K.~Abeling, D.K.~Abhayasinghe
  et~al., \emph{Search for invisible higgs-boson decays in events with
  vector-boson fusion signatures using 139 fb-1 of proton-proton data recorded
  by the atlas experiment},
  \href{https://doi.org/10.1007/jhep08(2022)104}{\emph{Journal of High Energy
  Physics} {\bfseries 2022} (2022) }.

\bibitem{CMS2PhysRevD.105.092007}
{\scshape CMS Collaboration} collaboration, \emph{Search for invisible decays
  of the higgs boson produced via vector boson fusion in proton-proton
  collisions at $\sqrt{s}=13\text{ }\text{ }\mathrm{TeV}$},
  \href{https://doi.org/10.1103/PhysRevD.105.092007}{\emph{Phys. Rev. D}
  {\bfseries 105} (2022) 092007}.

\bibitem{PhysRev.51.125}
W.H.~Furry, \emph{A symmetry theorem in the positron theory},
  \href{https://doi.org/10.1103/PhysRev.51.125}{\emph{Phys. Rev.} {\bfseries
  51} (1937) 125}.

\bibitem{Sakurai:2022tbk}
M.~Sakurai et~al., \emph{{muEDM: Towards a Search for the Muon Electric Dipole
  Moment at PSI Using the Frozen-spin Technique}},
  \href{https://doi.org/10.7566/JPSCP.37.020604}{\emph{JPS Conf. Proc.}
  {\bfseries 37} (2022) 020604}
  [\href{https://arxiv.org/abs/2201.06561}{{\ttfamily 2201.06561}}].

\bibitem{E137_Zboson_Bjorken:1988as}
J.D.~Bjorken, S.~Ecklund, W.R.~Nelson, A.~Abashian, C.~Church, B.~Lu et~al.,
  \emph{{Search for Neutral Metastable Penetrating Particles Produced in the
  SLAC Beam Dump}}, \href{https://doi.org/10.1103/PhysRevD.38.3375}{\emph{Phys.
  Rev. D} {\bfseries 38} (1988) 3375}.

\bibitem{E137_Zboson_Batell:2014mga}
B.~Batell, R.~Essig and Z.~Surujon, \emph{{Strong Constraints on Sub-GeV Dark
  Sectors from SLAC Beam Dump E137}},
  \href{https://doi.org/10.1103/PhysRevLett.113.171802}{\emph{Phys. Rev. Lett.}
  {\bfseries 113} (2014) 171802}
  [\href{https://arxiv.org/abs/1406.2698}{{\ttfamily 1406.2698}}].

\bibitem{E137_Zboson_Marsicano_2018}
L.~Marsicano, M.~Battaglieri, M.~Bondí, C.~Carvajal, A.~Celentano,
  M.~De~Napoli et~al., \emph{Dark photon production through positron
  annihilation in beam-dump experiments},
  \href{https://doi.org/10.1103/physrevd.98.015031}{\emph{Physical Review D}
  {\bfseries 98} (2018) }.

\bibitem{E141_ZbosonRiordan:1987aw}
E.M.~Riordan et~al., \emph{{A Search for Short Lived Axions in an Electron Beam
  Dump Experiment}},
  \href{https://doi.org/10.1103/PhysRevLett.59.755}{\emph{Phys. Rev. Lett.}
  {\bfseries 59} (1987) 755}.

\bibitem{CHARM_Zboson_Gninenko_2012}
S.~Gninenko, \emph{Constraints on sub-gev hidden sector gauge bosons from a
  search for heavy neutrino decays},
  \href{https://doi.org/10.1016/j.physletb.2012.06.002}{\emph{Physics Letters
  B} {\bfseries 713} (2012) 244–248}.

\bibitem{nucal_Zboson_Bl_mlein_2014}
J.~Blümlein and J.~Brunner, \emph{New exclusion limits on dark gauge forces
  from proton bremsstrahlung in beam-dump data},
  \href{https://doi.org/10.1016/j.physletb.2014.02.029}{\emph{Physics Letters
  B} {\bfseries 731} (2014) 320–326}.

\bibitem{Supernova_Zboson_Chang_2017}
J.H.~Chang, R.~Essig and S.D.~McDermott, \emph{Revisiting supernova 1987a
  constraints on dark photons},
  \href{https://doi.org/10.1007/jhep01(2017)107}{\emph{Journal of High Energy
  Physics} {\bfseries 2017} (2017) }.

\bibitem{LHCb_ZbosoPhysRevLett.124.041801}
{\scshape LHCb Collaboration} collaboration, \emph{Search for
  ${A}^{\ensuremath{'}}\ensuremath{\rightarrow}{\ensuremath{\mu}}^{+}{\ensuremath{\mu}}^{\ensuremath{-}}$
  decays}, \href{https://doi.org/10.1103/PhysRevLett.124.041801}{\emph{Phys.
  Rev. Lett.} {\bfseries 124} (2020) 041801}.

\bibitem{CMS_Zboson:2019kiy}
{\scshape CMS} collaboration, \emph{{Search for a narrow resonance decaying to
  a pair of muons in proton-proton collisions at 13 TeV}}, .

\bibitem{BaBar_Zboson:2014zli}
{\scshape BaBar} collaboration, \emph{{Search for a Dark Photon in $e^+e^-$
  Collisions at BaBar}},
  \href{https://doi.org/10.1103/PhysRevLett.113.201801}{\emph{Phys. Rev. Lett.}
  {\bfseries 113} (2014) 201801}
  [\href{https://arxiv.org/abs/1406.2980}{{\ttfamily 1406.2980}}].

\bibitem{NA482_Zboson:2015wmo}
{\scshape NA48/2} collaboration, \emph{{Search for the dark photon in $\pi^0$
  decays}}, \href{https://doi.org/10.1016/j.physletb.2015.04.068}{\emph{Phys.
  Lett. B} {\bfseries 746} (2015) 178}
  [\href{https://arxiv.org/abs/1504.00607}{{\ttfamily 1504.00607}}].

\bibitem{NA64_Zboson_Banerjee_2018}
D.~Banerjee, V.~Burtsev, A.~Chumakov, D.~Cooke, P.~Crivelli, E.~Depero et~al.,
  \emph{Search for a hypothetical 16.7 mev gauge boson and dark photons in the
  na64 experiment at cern},
  \href{https://doi.org/10.1103/physrevlett.120.231802}{\emph{Physical Review
  Letters} {\bfseries 120} (2018) }.

\bibitem{g_2e_Zboson_Pospelov_2009}
M.~Pospelov, \emph{Secluded u(1) below the weak scale},
  \href{https://doi.org/10.1103/physrevd.80.095002}{\emph{Physical Review D}
  {\bfseries 80} (2009) }.

\bibitem{eg-2Morel:2020dww}
L.~Morel, Z.~Yao, P.~Clad\'e and S.~Guellati-Kh\'elifa, \emph{{Determination of
  the fine-structure constant with an accuracy of 81 parts per trillion}},
  \href{https://doi.org/10.1038/s41586-020-2964-7}{\emph{Nature} {\bfseries
  588} (2020) 61}.

\bibitem{PAN2020114968}
J.-X.~Pan, M.~He, X.-G.~He and G.~Li, \emph{Scrutinizing a massless dark
  photon: Basis independence},
  \href{https://doi.org/https://doi.org/10.1016/j.nuclphysb.2020.114968}{\emph{Nuclear
  Physics B} {\bfseries 953} (2020) 114968}.

\bibitem{McKeen_2012}
D.~McKeen, M.~Pospelov and A.~Ritz, \emph{Modified higgs branching ratios
  versus cp and lepton flavor violation},
  \href{https://doi.org/10.1103/physrevd.86.113004}{\emph{Physical Review D}
  {\bfseries 86} (2012) }.

\bibitem{mQ6Davidson_2000}
S.~Davidson, S.~Hannestad and G.~Raffelt, \emph{Updated bounds on milli-charged
  particles},
  \href{https://doi.org/10.1088/1126-6708/2000/05/003}{\emph{Journal of High
  Energy Physics} {\bfseries 2000} (2000) 003–003}.

\bibitem{mQ1Vogel_2014}
H.~Vogel and J.~Redondo, \emph{Dark radiation constraints on minicharged
  particles in models with a hidden photon},
  \href{https://doi.org/10.1088/1475-7516/2014/02/029}{\emph{Journal of
  Cosmology and Astroparticle Physics} {\bfseries 2014} (2014) 029–029}.

\bibitem{mQ2Chang_2018}
J.H.~Chang, R.~Essig and S.D.~McDermott, \emph{Supernova 1987a constraints on
  sub-gev dark sectors, millicharged particles, the qcd axion, and an
  axion-like particle},
  \href{https://doi.org/10.1007/jhep09(2018)051}{\emph{Journal of High Energy
  Physics} {\bfseries 2018} (2018) }.

\bibitem{mQ3Badertscher_2007}
A.~Badertscher, P.~Crivelli, W.~Fetscher, U.~Gendotti, S.N.~Gninenko,
  V.~Postoev et~al., \emph{Improved limit on invisible decays of positronium},
  \href{https://doi.org/10.1103/physrevd.75.032004}{\emph{Physical Review D}
  {\bfseries 75} (2007) }.

\bibitem{mQ4Prinz_1998}
A.A.~Prinz, R.~Baggs, J.~Ballam, S.~Ecklund, C.~Fertig, J.A.~Jaros et~al.,
  \emph{Search for millicharged particles at slac},
  \href{https://doi.org/10.1103/physrevlett.81.1175}{\emph{Physical Review
  Letters} {\bfseries 81} (1998) 1175–1178}.

\bibitem{mQ5Magill_2019}
G.~Magill, R.~Plestid, M.~Pospelov and Y.-D.~Tsai, \emph{Millicharged particles
  in neutrino experiments},
  \href{https://doi.org/10.1103/physrevlett.122.071801}{\emph{Physical Review
  Letters} {\bfseries 122} (2019) }.

\bibitem{mQ7Jaeckel_2013}
J.~Jaeckel, M.~Jankowiak and M.~Spannowsky, \emph{Lhc probes the hidden
  sector}, \href{https://doi.org/10.1016/j.dark.2013.06.001}{\emph{Physics of
  the Dark Universe} {\bfseries 2} (2013) 111–117}.

\bibitem{mQ8PhysRevD.43.2314}
S.~Davidson, B.~Campbell and D.~Bailey, \emph{Limits on particles of small
  electric charge}, \href{https://doi.org/10.1103/PhysRevD.43.2314}{\emph{Phys.
  Rev. D} {\bfseries 43} (1991) 2314}.

\bibitem{Kalliokoski:2023cgw}
M.~Kalliokoski, V.A.~Mitsou, M.~de~Montigny, A.~Mukhopadhyay, P.-P.A.~Ouimet,
  J.~Pinfold et~al., \emph{{Searching for minicharged particles at the energy
  frontier with the MoEDAL-MAPP experiment at the LHC}},
  \href{https://doi.org/10.1007/JHEP04(2024)137}{\emph{JHEP} {\bfseries 04}
  (2024) 137} [\href{https://arxiv.org/abs/2311.02185}{{\ttfamily
  2311.02185}}].

\bibitem{WMAP_Dubovsky_2004}
S.L.~Dubovsky, D.S.~Gorbunov and G.I.~Rubtsov, \emph{Narrowing the window for
  millicharged particles by cmb anisotropy},
  \href{https://doi.org/10.1134/1.1675909}{\emph{Journal of Experimental and
  Theoretical Physics Letters} {\bfseries 79} (2004) 1–5}.

\bibitem{McDermott:2010pa}
S.D.~McDermott, H.-B.~Yu and K.M.~Zurek, \emph{{Turning off the Lights: How
  Dark is Dark Matter?}},
  \href{https://doi.org/10.1103/PhysRevD.83.063509}{\emph{Phys. Rev. D}
  {\bfseries 83} (2011) 063509}
  [\href{https://arxiv.org/abs/1011.2907}{{\ttfamily 1011.2907}}].

\bibitem{Bowman:2018yin}
J.D.~Bowman, A.E.E.~Rogers, R.A.~Monsalve, T.J.~Mozdzen and N.~Mahesh,
  \emph{{An absorption profile centred at 78 megahertz in the sky-averaged
  spectrum}}, \href{https://doi.org/10.1038/nature25792}{\emph{Nature}
  {\bfseries 555} (2018) 67}
  [\href{https://arxiv.org/abs/1810.05912}{{\ttfamily 1810.05912}}].

\bibitem{Liu:2019knx}
H.~Liu, N.J.~Outmezguine, D.~Redigolo and T.~Volansky, \emph{{Reviving
  Millicharged Dark Matter for 21-cm Cosmology}},
  \href{https://doi.org/10.1103/PhysRevD.100.123011}{\emph{Phys. Rev. D}
  {\bfseries 100} (2019) 123011}
  [\href{https://arxiv.org/abs/1908.06986}{{\ttfamily 1908.06986}}].

\bibitem{Aboubrahim:2021ohe}
A.~Aboubrahim, P.~Nath and Z.-Y.~Wang, \emph{{A cosmologically consistent
  millicharged dark matter solution to the EDGES anomaly of possible string
  theory origin}}, \href{https://doi.org/10.1007/JHEP12(2021)148}{\emph{JHEP}
  {\bfseries 12} (2021) 148}
  [\href{https://arxiv.org/abs/2108.05819}{{\ttfamily 2108.05819}}].

\bibitem{Munoz:2018jwq}
J.B.~Mu\~noz, C.~Dvorkin and A.~Loeb, \emph{{21-cm Fluctuations from Charged
  Dark Matter}},
  \href{https://doi.org/10.1103/PhysRevLett.121.121301}{\emph{Phys. Rev. Lett.}
  {\bfseries 121} (2018) 121301}
  [\href{https://arxiv.org/abs/1804.01092}{{\ttfamily 1804.01092}}].

\bibitem{Mathur:2021gej}
A.~Mathur, S.~Rajendran and H.~Ramani, \emph{{Composite solution to the EDGES
  anomaly}}, \href{https://doi.org/10.1103/PhysRevD.105.075020}{\emph{Phys.
  Rev. D} {\bfseries 105} (2022) 075020}
  [\href{https://arxiv.org/abs/2102.11284}{{\ttfamily 2102.11284}}].

\bibitem{tHooft:1974kcl}
G.~'t~Hooft, \emph{{Magnetic Monopoles in Unified Gauge Theories}},
  \href{https://doi.org/10.1016/0550-3213(74)90486-6}{\emph{Nucl. Phys. B}
  {\bfseries 79} (1974) 276}.

\bibitem{Polyakov:1974ek}
A.M.~Polyakov, \emph{{Particle Spectrum in Quantum Field Theory}}, {\emph{JETP
  Lett.} {\bfseries 20} (1974) 194}.

\bibitem{Brummer:2009cs}
F.~Brummer and J.~Jaeckel, \emph{{Minicharges and Magnetic Monopoles}},
  \href{https://doi.org/10.1016/j.physletb.2009.04.041}{\emph{Phys. Lett. B}
  {\bfseries 675} (2009) 360}
  [\href{https://arxiv.org/abs/0902.3615}{{\ttfamily 0902.3615}}].

\bibitem{Witten:1979ey}
E.~Witten, \emph{{Dyons of Charge e theta/2 pi}},
  \href{https://doi.org/10.1016/0370-2693(79)90838-4}{\emph{Phys. Lett. B}
  {\bfseries 86} (1979) 283}.

\bibitem{Brummer:2009oul}
F.~Brummer, J.~Jaeckel and V.V.~Khoze, \emph{{Magnetic Mixing: Electric
  Minicharges from Magnetic Monopoles}},
  \href{https://doi.org/10.1088/1126-6708/2009/06/037}{\emph{JHEP} {\bfseries
  06} (2009) 037} [\href{https://arxiv.org/abs/0905.0633}{{\ttfamily
  0905.0633}}].

\bibitem{MoEDAL:2020pyb}
{\scshape MoEDAL} collaboration, \emph{{First Search for Dyons with the Full
  MoEDAL Trapping Detector in 13 TeV $pp$ Collisions}},
  \href{https://doi.org/10.1103/PhysRevLett.126.071801}{\emph{Phys. Rev. Lett.}
  {\bfseries 126} (2021) 071801}
  [\href{https://arxiv.org/abs/2002.00861}{{\ttfamily 2002.00861}}].

\bibitem{bauer2018introductiondarkmatter}
M.~Bauer and T.~Plehn, \emph{Yet another introduction to dark matter},  2018.

\bibitem{Planck:2018vyg}
{\scshape Planck} collaboration, \emph{{Planck 2018 results. VI. Cosmological
  parameters}},
  \href{https://doi.org/10.1051/0004-6361/201833910}{\emph{Astron. Astrophys.}
  {\bfseries 641} (2020) A6}
  [\href{https://arxiv.org/abs/1807.06209}{{\ttfamily 1807.06209}}].

\bibitem{LUXdarkmatter}
D.~Akerib, S.~Alsum, H.~Araújo, X.~Bai, A.~Bailey, J.~Balajthy et~al.,
  \emph{Results from a search for dark matter in the complete lux exposure},
  \href{https://doi.org/10.1103/physrevlett.118.021303}{\emph{Physical Review
  Letters} {\bfseries 118} (2017) }.

\bibitem{XenonDirect}
{\scshape XENON Collaboration 7} collaboration, \emph{Dark matter search
  results from a one ton-year exposure of xenon1t},
  \href{https://doi.org/10.1103/PhysRevLett.121.111302}{\emph{Phys. Rev. Lett.}
  {\bfseries 121} (2018) 111302}.

\bibitem{PandaX-4TDarkMatter}
Y.~Meng, Z.~Wang, Y.~Tao, A.~Abdukerim, Z.~Bo, W.~Chen et~al., \emph{Dark
  matter search results from the pandax-4t commissioning run},
  \href{https://doi.org/10.1103/physrevlett.127.261802}{\emph{Physical Review
  Letters} {\bfseries 127} (2021) }.

\bibitem{LZDarkMatter}
{\scshape LUX-ZEPLIN Collaboration} collaboration, \emph{First dark matter
  search results from the lux-zeplin (lz) experiment},
  \href{https://doi.org/10.1103/PhysRevLett.131.041002}{\emph{Phys. Rev. Lett.}
  {\bfseries 131} (2023) 041002}.

\bibitem{Smith_2001}
D.~Smith and N.~Weiner, \emph{Inelastic dark matter},
  \href{https://doi.org/10.1103/physrevd.64.043502}{\emph{Physical Review D}
  {\bfseries 64} (2001) }.

\bibitem{Abdelhameed_2019}
A.~Abdelhameed, G.~Angloher, P.~Bauer, A.~Bento, E.~Bertoldo, C.~Bucci et~al.,
  \emph{First results from the cresst-iii low-mass dark matter program},
  \href{https://doi.org/10.1103/physrevd.100.102002}{\emph{Physical Review D}
  {\bfseries 100} (2019) }.

\bibitem{Aprile_2019}
E.~Aprile, J.~Aalbers, F.~Agostini, M.~Alfonsi, L.~Althueser, F.~Amaro et~al.,
  \emph{Light dark matter search with ionization signals in xenon1t},
  \href{https://doi.org/10.1103/physrevlett.123.251801}{\emph{Physical Review
  Letters} {\bfseries 123} (2019) }.

\bibitem{Bell_2021}
N.F.~Bell, J.B.~Dent, B.~Dutta, S.~Ghosh, J.~Kumar, J.L.~Newstead et~al.,
  \emph{Cosmic-ray upscattered inelastic dark matter},
  \href{https://doi.org/10.1103/physrevd.104.076020}{\emph{Physical Review D}
  {\bfseries 104} (2021) }.

\bibitem{Batell_2009}
B.~Batell, M.~Pospelov and A.~Ritz, \emph{Direct detection of multicomponent
  secluded wimps},
  \href{https://doi.org/10.1103/physrevd.79.115019}{\emph{Physical Review D}
  {\bfseries 79} (2009) }.

\bibitem{Griest:1990kh}
K.~Griest and D.~Seckel, \emph{{Three exceptions in the calculation of relic
  abundances}}, \href{https://doi.org/10.1103/PhysRevD.43.3191}{\emph{Phys.
  Rev. D} {\bfseries 43} (1991) 3191}.

\bibitem{ParticleDataGroup:2020ssz}
{\scshape Particle Data Group} collaboration, \emph{{Review of Particle
  Physics}}, \href{https://doi.org/10.1093/ptep/ptaa104}{\emph{PTEP} {\bfseries
  2020} (2020) 083C01}.

\bibitem{abdullahi2023semivisibledarkphotonphenomenology}
A.M.~Abdullahi, M.~Hostert, D.~Massaro and S.~Pascoli, \emph{Semi-visible dark
  photon phenomenology at the gev scale},  2023.

\bibitem{Mohlabeng_2019}
G.~Mohlabeng, \emph{Revisiting the dark photon explanation of the muon
  anomalous magnetic moment},
  \href{https://doi.org/10.1103/physrevd.99.115001}{\emph{Physical Review D}
  {\bfseries 99} (2019) }.

\bibitem{Izaguirre_2016}
E.~Izaguirre, G.~Krnjaic and B.~Shuve, \emph{Discovering inelastic thermal
  relic dark matter at colliders},
  \href{https://doi.org/10.1103/physrevd.93.063523}{\emph{Physical Review D}
  {\bfseries 93} (2016) }.

\bibitem{Izaguirre_2017}
E.~Izaguirre, Y.~Kahn, G.~Krnjaic and M.~Moschella, \emph{Testing light dark
  matter coannihilation with fixed-target experiments},
  \href{https://doi.org/10.1103/physrevd.96.055007}{\emph{Physical Review D}
  {\bfseries 96} (2017) }.

\bibitem{BaBar_invisible_Lees_2017}
J.~Lees, V.~Poireau, V.~Tisserand, E.~Grauges, A.~Palano, G.~Eigen et~al.,
  \emph{Search for invisible decays of a dark photon produced in $e^{+}e^{-}$
  collisions at babar},
  \href{https://doi.org/10.1103/physrevlett.119.131804}{\emph{Physical Review
  Letters} {\bfseries 119} (2017) }.

\bibitem{BESIII_Ablikim_2023}
M.~Ablikim, M.~Achasov, P.~Adlarson, M.~Albrecht, R.~Aliberti, A.~Amoroso
  et~al., \emph{Search for invisible decays of a dark photon using $e^{+}e^{-}$
  annihilation data at besiii},
  \href{https://doi.org/10.1016/j.physletb.2023.137785}{\emph{Physics Letters
  B} {\bfseries 839} (2023) 137785}.

\bibitem{NA64_invisible_Andreev_2021}
Y.~Andreev, D.~Banerjee, J.~Bernhard, M.~Bondì, V.~Burtsev, A.~Celentano
  et~al., \emph{Improved exclusion limit for light dark matter from
  $e^{+}e^{-}$ annihilation in na64},
  \href{https://doi.org/10.1103/physrevd.104.l091701}{\emph{Physical Review D}
  {\bfseries 104} (2021) }.

\bibitem{NA64:2016oww}
{\scshape NA64} collaboration, \emph{{Search for invisible decays of sub-GeV
  dark photons in missing-energy events at the CERN SPS}},
  \href{https://doi.org/10.1103/PhysRevLett.118.011802}{\emph{Phys. Rev. Lett.}
  {\bfseries 118} (2017) 011802}
  [\href{https://arxiv.org/abs/1610.02988}{{\ttfamily 1610.02988}}].

\bibitem{NA62:2019meo}
{\scshape NA62} collaboration, \emph{{Search for production of an invisible
  dark photon in $\pi^0$ decays}},
  \href{https://doi.org/10.1007/JHEP05(2019)182}{\emph{JHEP} {\bfseries 05}
  (2019) 182} [\href{https://arxiv.org/abs/1903.08767}{{\ttfamily
  1903.08767}}].

\bibitem{SMpredictioneEDM}
Y.~Ema, T.~Gao and M.~Pospelov, \emph{Standard model prediction for
  paramagnetic electric dipole moments},
  \href{https://doi.org/10.1103/PhysRevLett.129.231801}{\emph{Phys. Rev. Lett.}
  {\bfseries 129} (2022) 231801}.

\bibitem{Botella:2004ks}
F.J.~Botella, M.~Nebot and O.~Vives, \emph{{Invariant approach to
  flavor-dependent CP-violating phases in the MSSM}},
  \href{https://doi.org/10.1088/1126-6708/2006/01/106}{\emph{JHEP} {\bfseries
  01} (2006) 106} [\href{https://arxiv.org/abs/hep-ph/0407349}{{\ttfamily
  hep-ph/0407349}}].

\bibitem{Davidson:2005cw}
S.~Davidson and H.E.~Haber, \emph{{Basis-independent methods for the
  two-Higgs-doublet model}},
  \href{https://doi.org/10.1103/PhysRevD.72.099902}{\emph{Phys. Rev. D}
  {\bfseries 72} (2005) 035004}
  [\href{https://arxiv.org/abs/hep-ph/0504050}{{\ttfamily hep-ph/0504050}}].

\bibitem{Chang:1993kw}
D.~Chang, W.S.~Hou and W.-Y.~Keung, \emph{{Two loop contributions of flavor
  changing neutral Higgs bosons to mu ---\ensuremath{>} e gamma}},
  \href{https://doi.org/10.1103/PhysRevD.48.217}{\emph{Phys. Rev. D} {\bfseries
  48} (1993) 217} [\href{https://arxiv.org/abs/hep-ph/9302267}{{\ttfamily
  hep-ph/9302267}}].

\end{thebibliography}\endgroup
\newpage
\bibliographystyle{JHEP}
\end{document}